\documentclass[preprint,12pt]{elsarticle}

\usepackage{amssymb}

\journal{Nuclear Physics A}

\begin{document}

\begin{frontmatter}



\title{Effects of the Coulomb interaction on parameters of resonance states in mirror
three-cluster nuclei}

\author[a1]{A. D. Duisenbay}
\ead{duisenbay.aknur@gmail.com}

\address[a1]{
 Al-Farabi Kazakh National University, \\
Almaty 050040, Kazakhstan
}
\author[a1]{N. Kalzhigitov}
\ead{knurto1@gmail.com}

\author[a2]{K. Kat\=o}
\ead{kato@nucl.sci.hokudai.ac.jp}
\address[a2]{
 Nuclear Reaction Data Centre, Faculty of Science, Hokkaido University,\\
Sapporo 060-0810, Japan
}

\author[a1]{V. O. Kurmangaliyeva}

\author[a1]{N. Takibayev}
\ead{takibayev@gmail.com}

\author[a3]{V. S. Vasilevsky\corref{cor1}}
\ead{vsvasilevsky@gmail.com}\cortext[cor1]{}

\address[a3]{Bogolyubov Institute for Theoretical Physics,\\
 Kiev 03143, Ukraine
}



\date{\today }

\begin{abstract}
We investigate how the Coulomb interaction affects the energy $E$ and width
$\Gamma$ of resonance states in mirror nuclei. We employ a three-cluster
microscopic model to determine position of resonance states in two- and
three-body continua. Two parameters are introduced to quantify effects of the
Coulomb interactions. As the energy and width of the corresponding resonance
states of mirror nuclei are displayed on an $E$-$\Gamma$ plane, these
parameters determine a rotation and a dilatation. With the help of these
parameters we found resonance states with strong, small and medium effects of
the Coulomb interaction. We also found two different scenarios of the motion
of resonance states due to the Coulomb interaction. The first standard (major)
scenario represent resonance states with the larger energy and larger width
than their counterparts have. The second rear scenario includes resonance
states with the larger energy but smaller width.

\end{abstract}




\end{frontmatter}



\section{Introduction}

The main aim of this paper is to study effects of the Coulomb forces on the
energy and width of resonance states residing in two- and three-cluster
continua. We believe that the ideal objects for these studies are mirror
nuclei. If we formulate our many-cluster model in such a way that
inter-cluster interactions, originated only from a nucleon-nucleon
interaction, are the same in both mirror nuclei, then the relative position of
bound and resonance states and their widths \ are totally determined by the
Coulomb interaction of protons. Consider, for example, the mirror nuclei
$^{8}$Li and $^{8}$B. It is naturally to present them as three-cluster
configurations $\alpha+t+n$ and $\alpha+^{3}$He$+p$, respectively. Cluster
models with such three-cluster configurations are shown repeatedly 
\cite{2000PhRvC..61b4311C, 1994NuPhA.577..624B, 1994NuPhA.567..341D, 
 1995NuPhA.588..157V, 1977NuPhA.289..317S, 
 2017UkrJPh..62..461V} to provide
the correct description of many observed properties of these nuclei. In the
nucleus $^{8}$Li, the Coulomb interaction affects the interaction between an
alpha particle and a triton only. In the mirror $^{8}$B nucleus, the Coulomb
interaction reduces the effective attraction in the all pairs of interacting
clusters: $\alpha+^{3}$He, $\alpha+p$ and $^{3}$He$+p$. Moreover, the Pauli
principle generates the three-body Coulomb interaction, provided that the full
antisymmetrization of a compound system is taken into account correctly.

This problem has been repeatedly studied in literature
\cite{1961PhRvL...6...17W, 1966PhL....19..676O,
1963NucPh..45..437F, 1967NuPhA..96..593H,
1970NuPhA.144..539E, 1971PhLB...35..135V,
1977NuPhA.281..267B, 1997PhLB..414...13A,
1998PThPh..99..623A, 2001PhRvC..64b4305A,
2002CzJPS..52C.597B, 2002CoTPh..37..327O,
2004PhRvC..69b4002G, 2006JPhCS..49...85L,
2007EPJA...34..315H, 2014PTEP.2014h3D01M,
2015PhRvC..91b4325G, 2016EPJWC.11706014I,
2017EPJWC.16300040N, 2018PhRvC..98e4318N,
PhysRevC.99.021303}. However, in many of these publications, the main
attention was devoted to bound states. Meanwhile the most intriguing is the
impact of the Coulomb interaction on resonance states. There are some new publications
 \cite{2019PhRvL.122l2501W,2019PhRvC..99e4302W} dealing with this problem. 
In Ref. \cite{2019PhRvL.122l2501W}, the structure of mirror nuclei $^{11}$Li 
and $^{11}$O has been studied experimentally and within the Gamow coupled-channel
approach \cite{2017PhRvC..96d4307W}. Within this approach, the mirror nuclei $^{11}$Li 
and $^{11}$O are considered as three-body system with an inert core and two valence 
neutrons and protons, respectively. The calculated density distributions explicitly 
demonstrate effects of the Coulomb interaction on structure of the  
$J^{\pi}$ =3/2$^{-}_{1}$, 3/2$^{-}_{2}$, 5/2$^{+}_{1}$ and 5/2$^{+}_{2}$ resonance states. 
In Ref. \cite{2019PhRvC..99e4302W} similar analysis was performed to study
mirror nuclei $^{11}$Li-$^{11}$O and $^{12}$Be-$^{12}$O.

Effects of the Coulomb interaction on mirror or isobaric nuclei have been
repeatedly investigated by many authors in different many-particle models.
Very often the influence of the Coulomb potential on the spectrum of such
nuclei is associated with the Thomas-Erhman effect or shift (see original
papers Ref. \cite{1952PhRv...88.1109T} and Ref. \cite{1951PhRv...81..412E}
\ and recent discussion of the effect, for example, in Ref.
\cite{2007EPJA...34..315H}), which is connected with the shift of energy of
single particle levels in mirror nuclei due to the Coulomb interaction.
Recently this effect is also discussed in context of a cluster model.

By analyzing the spectra of the mirror nuclei $^{13}$C and $^{13}$N, Thomas in
Ref. \cite{1952PhRv...88.1109T} and Erhman in Ref. \cite{1951PhRv...81..412E}
independently discovered that the almost degenerated single-particle $s$- and
$d$-orbitals give the different contribution to the spectrum of a compound
nucleus due to the Coulomb interaction. The more compact orbital yields the
larger Coulomb shift of the single-particle energy than the more dispersed
orbital. Such a difference in contribution of the Coulomb forces with compact
and dispersed single-particle orbitals is called the Thomas-Erhman effect.
Since these publications, the Thomas-Erhman effect has been numerously
examined in different mirror nuclei. Last decades, this effect is intensively
studied within many-cluster models (see, for example, Refs.
\cite{1997PhLB..414...13A, 1998PThPh..99..623A,
2015PhRvC..91b4325G, 2016EPJWC.11706014I,
2018PhRvC..98e4318N, PhysRevC.99.021303}). It was shown that the
different cluster orbitals utilizing for a description of mirror nuclei give
also different contribution of the Coulomb energy. Some of these orbitals
describes relatively compact many-cluster configurations, and other orbitals
suggest the loosely many-cluster configurations. For example, in Ref.
\cite{PhysRevC.99.021303} the mirror nuclei $^{14}$C and $^{14}$O have been
studied with the antisymmetric molecular dynamics (AMD) and it was
demonstrated that the Coulomb potential had a different contribution to the
triangular, and linear $\sigma$- and $\pi$-bond configurations. 

We will not discuss the Thomas-Erhman effect as this is out of the scope 
of the present paper. Such a discussion requires a decomposing of complicated 
 wave functions of the three-clusters systems into simple orbitals. And it leads to very bulky calculations. 
Below, we will study the properties 
wave functions of bound and resonance states and employ other way of their decomposition which is used in many-cluster models. The main aim of the present paper is to 
suggest an adequate way (manner) of analysis of resonance state behavior 
in mirror nuclei and to apply it for reveal  general features of motion 
of resonance states in real three-cluster systems caused by the Coulomb forces.

In Ref. \cite{2017PhRvC..96c4322V} the impact of the Coulomb interaction on
energy and width of resonance states in three-cluster continua $\alpha
+\alpha+n$ and $\alpha+\alpha+p$ \ of the mirror nuclei $^{9}$Be and $^{9}$B
have been studied. As resonance states being poles of the $S$ matrix in the
complex plane, it was introduced a Coulomb rotating angle to determine and
to quantify how strong are the effects of Coulomb interactions. With this
parameter, it was discovered three groups of resonance states with the weak,
medium and strong impact of the interaction on the position of resonance
states. However, we feel that this analysis was not complete. To make this
analysis more complete we introduce a new parameter which determines the
relative shift of the energy and width of the resonance state in a mirror nucleus
with a large number of protons due to the stronger Coulomb interaction.

We are going to perform such an analysis for different couples of mirror nuclei,
namely, $^{7}$Li and $^{7}$Be, $^{8}$Li and $^{8}$B, $^{11}$B and $^{11}$C.
All these nuclei are considered within a three-cluster microscopic model. For
all these nuclei we selected dominant three-cluster configurations. In Table
\ref{Tab:MirrorNuclei} we show
partners of mirror nuclei and their dominant three-cluster channels. The
partners are marked by the letters $L$ and $R$. In Table
\ref{Tab:MirrorNuclei} we also show a microscopic model applied and a source
of calculations, and the charge difference $\Delta Z=Z_{R}-Z_{L}$ as well.%

\begin{table}[htbp] \centering
\caption{List of nuclei to be investigated, dominant three-cluster configurations, 
a microscopic model applied  and references.}%
\begin{tabular}
[c]{ccccc}\hline
$L$-nucleus & $R$-nucleus & $\Delta Z$ & Source & Model\\\hline
$^{7}$Li=$\alpha+d+n$ & $^{7}$Be=$\alpha+d+p$ & 1 & \cite{2009PAN....72.1450N,%
 2009NuPhA.824...37V} & AMGOB\\
$^{8}$Li=$\alpha+t+n$ & $^{8}$B=$\alpha+^{3}$He$+p$ & 2 &
\cite{2017UkrJPh..62..461V} & AMGOB\\
$^{9}$Be=$\alpha+\alpha+n$ & $^{9}$B=$\alpha+\alpha+p$ & 1 &
\cite{2017PhRvC..96c4322V, 2014PAN..77.555N} & AMHHB\\
$^{11}$B=$\alpha+\alpha+t$ & $^{11}$C=$\alpha+\alpha+^{3}$He & 1 &
\cite{2013UkrJPh.58.544V} & AMHHB\\\hline
\end{tabular}
\label{Tab:MirrorNuclei}%
\end{table}%

To study effects of the Coulomb interaction on resonance states in
three-cluster systems, we employ two microscopical models as shown in Table 
\ref{Tab:MirrorNuclei}. They are a
modification of the resonating group method. These methods are designed to
study a three-cluster structure of light atomic nuclei. Both of these methods
employ the square-integrable bases to describe dynamics of inter-cluster
motion. The first model was formulated in Ref. \cite{2009NuPhA.824...37V} and
will be referred as AMGOB, it utilizes of the Gaussian basis to describe bound
and pseudo-bound states in a two-cluster subsystem, while Oscillator basis
describes realative motion of the third cluster with two-cluster subsystem. 
The main merit of the AMGOB is that it allows us to study influence of a
cluster polarization, i.e. to study how the shape and size of a nucleus
comprised of two clusters are changed when another nucleus (the third
cluster) is moving closer. Thus this model provides more exact description
of nuclei \ with prominent two-cluster structure. By exploiting a
three-cluster configuration, the method involves up to three different
binary channels with the lowest energy of two-body decay of a compound
three-cluster nucleus. It was shown in Refs. \cite{2009NuPhA.824...37V, 
2009PAN....72.1450N, 2017UkrJPh..62..461V, 2012PAN.75.818V}
that the cluster polarization has a large impact on the spectrum of bound
and resonance states of light nuclei, and especially on the astrophysical
$S$-factor of the capture reactions.
The second model (AMHHB) is used the hyperspherical harmonics to investigate
relative motions of clusters and was designed in Ref.
\cite{2001PhRvC..63c4606V} to study processes in the three-cluster continuum.

The layout of our paper is following. In Sec. \ref{Sec:Coulomb}, we formulate in more detail 
our main aims and consider possible scenarios of motion of resonance states in mirror nuclei 
stimulated by the Coulomb forces. In Sec. \ref{Sec:Coulomb}  we also introduce
parameters which allow us to study thoroughly effects of the Coulomb
interaction on bound and resonance states. A short explanation of the main idea of the microscopic
method 
for description of bound and scattering states of
a three-cluster system are presented in Sec. \ref{Sec:Model}. In what follows, 
we will concentrate our main attention on resonance states in three-cluster 
continuum of mirror nuclei. That is why in Sec. \ref{Sec:Model} we present only 
one microscopic model (AMHHB) of two mentioned above.  How strong are the effects of the
Coulomb interaction are demonstrated in Sec. \ref{Sec:Effects}. We start by
considering resonance states in the three-cluster continuum of the mirror
nuclei $^{9}$Be and $^{9}$B, $^{11}$B and $^{11}$C, and then we proceed with
bound and resonance states in the two-body continuum of $^{7}$Li and $^{7}$Be,
$^{8}$Li and $^{8}$B. Analysis of wave functions of resonance states in the mirror nuclei is presented 
in Sec. \ref{Sec:WaveFuns}. We close the paper by summarizing the obtained results
in Sec. \ref{Sec:Conclusions}.

\section{The Coulomb interaction in mirror nuclei \label{Sec:Coulomb}}

\begin{figure}
[ptbh]
\begin{center}
\includegraphics[width=\columnwidth]
{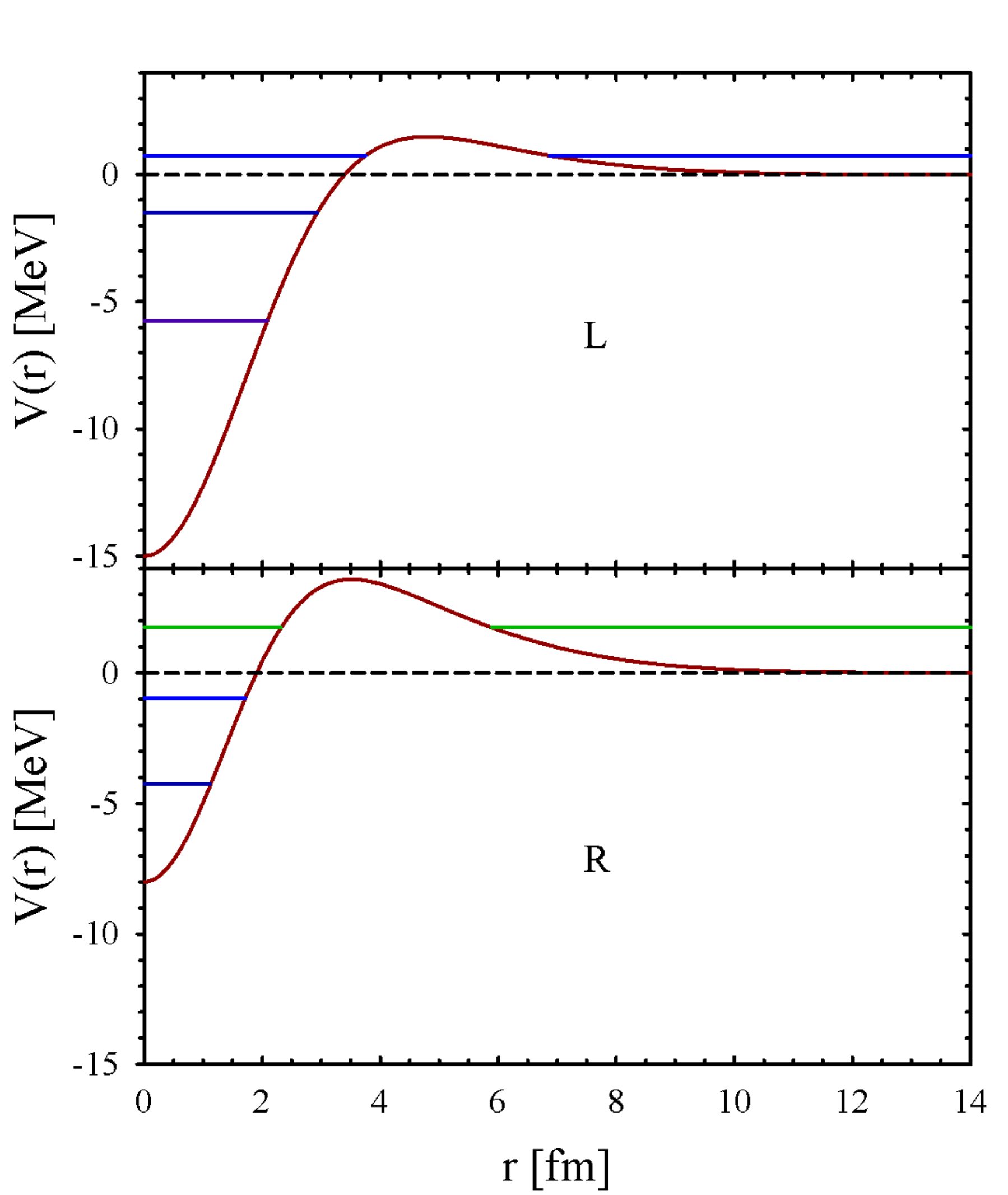}%
\caption{Effective potentials $V\left(  r\right)  $ of two mirror nuclei which have the charges $Z_L<Z_R$ as a
function of distance $r$.}%
\label{Fig:EffPotentMN}%
\end{center}
\end{figure}

To formulate more clearly our aim, let us consider a schematic picture which
demonstrates effects of the Coulomb interaction in mirror nuclei. By letters
$L$ and $R$ we denote two mirror nuclei assuming that the charges of these
nuclei obey the relation \ $Z_{L}<Z_{R}$. \ In Fig. \ref{Fig:EffPotentMN} we
show effective potentials of two mirror nuclei and the position of two bound
states and one resonance state.%

The next figure (Fig. \ref{Fig:EffPotentS}) demonstrates the effective
potentials above the decay threshold. This picture shows that the Coulomb
interaction increases the height and width of the effective barrier. This
figure also suggests the two possible scenarios representing effects of the
Coulomb interaction on parameters of resonance states. The first scenario
assumes that the Coulomb interaction increases energy of the resonance state
in the $R$ nucleus in such a way that the width of the Coulomb barrier at this
energy becomes rather small, which results in increasing of the resonance
width. In the second scenario the energy of a resonance state in the $R$
nucleus is also increased but not so high as in the first scenario. At this
energy the width of the Coulomb barrier is large and yields a smaller width
of the resonance state.%

\begin{figure}
[ptbh]
\begin{center}
\includegraphics[width=\columnwidth]
{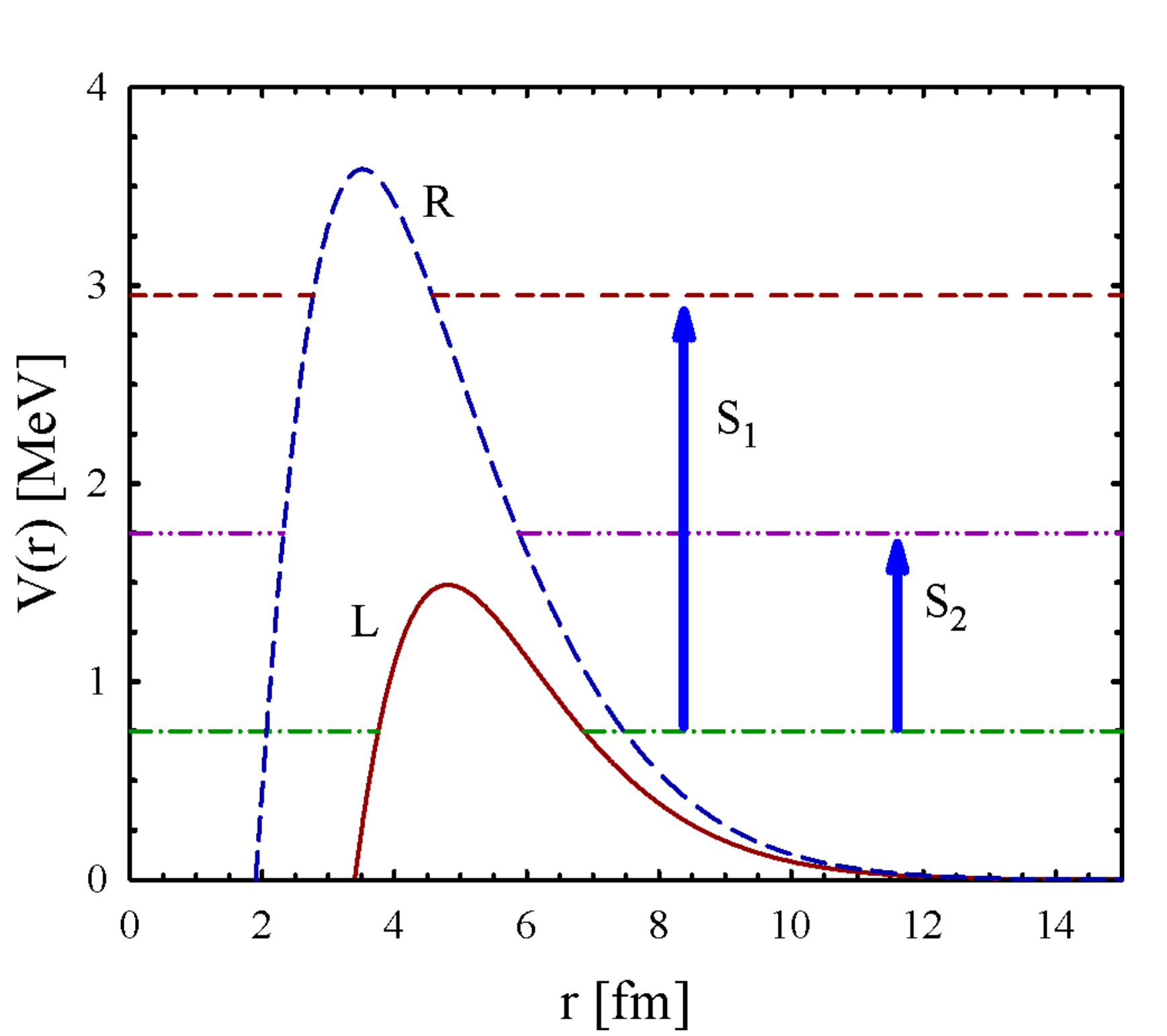}%
\caption{Effective potential barriers in mirror nuclei and position of
resonance states.}%
\label{Fig:EffPotentS}%
\end{center}
\end{figure}

By considering the mirror nuclei, one can suggest four main scenarios for
changing of the parameters of resonance states due to the Coulomb forces.
Increasing of the Coulomb interaction leads to decreasing of the attractive
effective interaction in each channel of a many-channel system. That may shift
up the energy of resonance states and may also increase the width of resonance
states. That is the first scenario. The second scenario,  the Coulomb
interaction makes an effective barrier more thicker, that may increase the
energy but decrease the width of a resonance state. We add the third and fourth
scenarios, when the more wider effective barrier may decrease the energy of 
resonance states and increase or decrease the width of resonance states, respectively. In  
the present paper we will investigate what
scenario dominates in three-cluster mirror nuclei.

Resonance states are characterized by two parameters: the energy $E$ and width
$\Gamma$. Being a pole of the $S$-matrix, the resonance state is usually
determined by a complex value $E-i\Gamma/2$. Thus, it is natural to consider
the parameters of the resonance state in a two-dimensional space. We select
the plane $E$ and $\Gamma$. In this plane the impact of the Coulomb forces on
the resonance state can be reduced to two operations: a rotation and a shift.
That is why we introduce the Coulomb rotational angle $\theta_{C}$
\begin{equation}
\theta_{C}=\arctan\left[  \frac{\Gamma\left(  R\right)  -\Gamma\left(
L\right)  }{E\left(  R\right)  -E\left(  L\right)  }\right]  \label{eq:C01}%
\end{equation}
and the Coulomb shift $R_{C}$%
\begin{equation}
R_{C}=\sqrt{\left[  E\left(  R\right)  -E\left(  L\right)  \right]
^{2}+\left[  \Gamma\left(  R\right)  -\Gamma\left(  L\right)  \right]  ^{2}%
}.\label{eq:C02}%
\end{equation}
These relations connect resonance states of mirror nuclei with the same total angular 
momentum $J$ and the same parity $\pi$. They can be also applied for a bound state in the $L$
nucleus and a resonance state in the $R$ nucleus, and for two bound
states in these nuclei. In the latter case,  these relations give the trivial result: $\theta_{C}=0$ and $R_{C}=E\left(
R\right)  -E\left(  L\right)  $.


The Coulomb rotational angle $\theta_{C}$ was introduced in Ref.
\cite{2017PhRvC..96c4322V} and used to study effects of the Coulomb forces in
$^{9}$Be and $^{9}$B.

Formulae (\ref{eq:C01}) and (\ref{eq:C02}) suggest that it is more expedient
to use differences
\begin{eqnarray}
\Delta E  &  =& E\left(  R\right)  -E\left(  L\right)  =R_{C}\cos\theta
_{C},\label{eq:C03}\\
\Delta\Gamma &  = & \Gamma\left(  R\right)  -\Gamma\left(  L\right)  =R_{C}%
\sin\theta_{C}\nonumber
\end{eqnarray}
as the axis $x$ and axis $y$, respectively. Both differences can be positive
or negative. In term of the parameters $\Delta E$ and $\Delta\Gamma$ we can
consider the four following hypothetical scenarios:

\begin{description}
\item[S$_{1}$] Both parameters are positive, thus the Coulomb interaction
increases the energy and width of resonance state in the $R$ nucleus.

\item[S$_{2}$] The parameter $\Delta E$ is positive and $\Delta\Gamma$ is
negative, the energy of resonance state in the $R$ nucleus is larger however
its width is smaller than in the $L$ nucleus.

\item[S$_{3}$] The parameter $\Delta E$ is negative and $\Delta\Gamma$ is
positive, the Coulomb interaction reduces the energy of the resonance state in
the $R$ nucleus but increases its width.

\item[S$_{4}$] Both parameters are negative, it means the Coulomb interaction
makes smaller the energy and width of the resonance state in the $R$ nucleus.
\end{description}

Note that in the first and second scenarios the Coulomb interaction increases
the energy of resonance state in $R$ nucleus, meanwhile the third and fourth
scenarios\ assume that the energy is diminished by the Coulomb interactions.

\section{Model formulation \label{Sec:Model}}

In this section we shortly present the main ideas of a microscopic
three-cluster model used to study spectrum of bound and resonance states in
mirror nuclei. The principal restrictions and approximations in any model are
imposed on a Hamiltonian and wave functions. In the present model we use a
microscopic Hamiltonian which includes the kinetic energy in the system of
center mass of $A$ nucleons, central and spin-orbital components of a
semirealistic nucleon-nucleon interaction and the Coulomb interaction of
protons. Exact wave function of the Hamiltonian is approximated by the wave
function for the system consisting of three $s$-clusters%
\begin{eqnarray}
\Psi_{JM_{J}}^{E}  & =&\sum_{L,S,\lambda,l}\widehat{\mathcal{A}}\left\{
\left[  \Phi_{1}\left(  A_{1},b,s_{1}\right)  \Phi_{2}\left(  A_{2}%
,b,s_{2}\right)  \Phi_{3}\left(  A_{3},b,s_{3}\right)  \right]  _{S}\right.
\label{eq:001}\\
& \times& \left.  \psi_{\lambda,l;LS}^{E,J}\left(  x,y\right)  \left\{
Y_{\lambda}\left(  \widehat{\mathbf{x}}\right)  Y_{l}\left(  \widehat
{\mathbf{y}}\right)  \right\}  _{L}\right\}  _{JM_{J}}.\nonumber
\end{eqnarray}
The internal structure of clusters ($\alpha$= 1, 2, 3) is described by the
antisymmetric and translationally invariant wave functions $\Phi_{\alpha
}\left(  A_{\alpha},b,s_{\alpha}\right)  $. In this function we indicated the
main parameters which determines this function: $A_{\alpha}$ is the number of
nucleons, $b$\ is an oscillator length and the spin $s_{\alpha}$ of the
cluster. The function $\Phi_{\alpha}\left(  A_{\alpha},b_{\alpha},s_{\alpha
}\right)  $ is a wave function of the many-particle shell model with the most
compact configuration of nucleons. The antisymmetrization operator
$\widehat{\mathcal{A}}$ makes antisymmetric the wave function of the compound
three-cluster system. Within the standard approximation of the resonating
group method all functions $\Phi_{\alpha}\left(  A_{\alpha},b,s_{\alpha
}\right)  $ are fixed, and thus to calculate a spectrum and wave functions of
the compound system one has to determine a wave function of inter-cluster
motion $\psi_{\lambda,l;LS}^{E,J}\left(  x,y\right)  $. This function is also
translationally invariant and depends on two Jacobi vectors $\mathbf{x}$ and
$\mathbf{y}$, locating relative position of clusters in the space. The vector
$\mathbf{x}$ determines the distance between selected pair of clusters and the
vector $\mathbf{y}$ is proportional to the displacement of the third cluster
with respect to the center of mas of two-cluster subsystem. In Eq.
(\ref{eq:001}) $\widehat{\mathbf{x}}$ and $\widehat{\mathbf{y}}$ are unit
vectors, and $\lambda$ and $l$ are the partial angular momenta associated with
the vectors $\mathbf{x}$ and $\mathbf{y}$ respectively.

Wave functions of inter-cluster motion $\psi_{\lambda,l;LS}^{E,J}\left(
x,y\right)  $ obey an infinite set of the two-dimensional (in terms of
variables $x\ $and $y$) integro-differential equations. Two-dimensional
equations can be reduced to one-dimensional integro-differential equations by
employing the hyperspherical coordinates (one hyper radius $\rho$ and five
hyperspherical angles $\Omega_{5}$) and hyperspherical harmonics.

Instead of six variables $\mathbf{x}$ and $\mathbf{y}$ or $x$, $y$, and two
unit vectors $\widehat{\mathbf{x}}$ and $\widehat{\mathbf{y}}$ we introduce a
hyperspherical radius $\rho$ and a hyperspherical angle $\theta$%
\begin{equation}
\rho=\sqrt{x^{2}+y^{2}},\quad\theta=\arctan\left(  \frac{x}{y}\right)  .
\label{eq:004}%
\end{equation}
For a fixed value of $\rho$, the angle $\theta$ determines relative length of
the vectors $\mathbf{x}$ and $\mathbf{y}$%
\begin{equation}
x=\rho\cos\theta,\quad y=\rho\sin\theta,\quad\theta\in\left[  0,\pi/2\right]
. \label{eq:005}%
\end{equation}
Thus the set of hyperspherical angles $\Omega_{5}$ is $\Omega_{5}=\left\{
\theta,\widehat{\mathbf{x}},\widehat{\mathbf{y}}\right\}  =\left\{
\theta,\theta_{x},\phi_{x},\theta_{y},\phi_{y}\right\}  $. With such a
definition of the hyperspherical angles, we can employ the hyperspherical
harmonics in the form suggested by Zernike and Brinkman in Ref. \cite{kn:ZB}.
They have simple form and the quantum numbers numerating them\ have clear
physical meaning.

The represented set of the hyperspherical angles is very popular scheme of the
hyperspherical coordinates for investigating three-body \cite{kn:Zhuk93E,
2001FBS....30...39V, 2001FBS....31...13C}, and three-cluster systems
\cite{2001PhRvC..63c4606V, 2001PhRvC..63c4607V, 2010PPN....41..716N,
2004NuPhA.740..249K}.

In new coordinates%
\begin{eqnarray}
\Psi_{JM_{J}}^{E}  & =&\sum_{L,S,\lambda,l}\widehat{\mathcal{A}}\left\{
\left[  \Phi_{1}\left(  A_{1},b_{1},s_{1}\right)  \Phi_{2}\left(  A_{2}%
,b_{2},s_{2}\right)  \Phi_{3}\left(  A_{3},b_{3},s_{3}\right)  \right]
_{S}\right.  \label{eq:006}\\
& \times& \left.  \phi_{c}^{E,J}\left(  \rho\right)  \mathcal{Y}_{c}\left(
\Omega_{5}\right)  \right\}  _{JM_{J}},\nonumber
\end{eqnarray}
where $\mathcal{Y}_{c}\left(  \Omega_{5}\right)  $ stands\ for the product
\begin{equation}
\mathcal{Y}_{c}\left(  \Omega_{5}\right)  =\chi_{K}^{\left(  \lambda,l\right)
}\left(  \theta\right)  \left\{  Y_{\lambda}\left(  \widehat{\mathbf{x}%
}\right)  Y_{l}\left(  \widehat{\mathbf{y}}\right)  \right\}  _{LM_{L}}
\label{eq:007}%
\end{equation}
and represents a hyperspherical harmonic for a three-cluster channel
\begin{equation}
c=\left\{  K,\lambda,l,L\right\}  . \label{eq:007A}%
\end{equation}
Definition of all components of the hyperspherical harmonic $Y_{c}\left(
\Omega_{5}\right)  $ can be found, for instance, in Ref.
\cite{2001PhRvC..63c4606V}. Being a complete basis, the hyperspherical
harmonics account for any shape of the three-cluster triangle and its
orientation and thus they allow one to describe all possible modes of relative
motion of \ three interacting clusters.

To determine the hyperradial wave functions $\phi_{c}^{E,J}\left(  \rho\right)  $ one has to solve a
system of integro--differential equations with nonlocal effective potentials
for three clusters. This system of equations can be represented as%
\begin{eqnarray}
& & \sum_{\widetilde{c}}\left[  \delta_{c,\widetilde{c}}\widehat{T}_{K}\phi
_{c}^{E,J}+\int d\widetilde{\rho}\widetilde{\rho}^{5}V_{c,\widetilde{c}%
}\left(  \rho,\widetilde{\rho}\right)  \phi_{\widetilde{c}}^{E,J}\left(
\widetilde{\rho}\right)  \right]  \label{eq:008}\\
& =&E\sum_{\widetilde{c}}\int d\widetilde{\rho}\widetilde{\rho}^{5}%
N_{c,\widetilde{c}}\left(  \rho,\widetilde{\rho}\right)  \phi_{\widetilde{c}%
}^{E,J}\left(  \widetilde{\rho}\right)  ,\nonumber
\end{eqnarray}
where%
\begin{equation}
\widehat{T}_{K}=-\frac{\hbar^{2}}{2m}\left[  \frac{\partial^{2}}{\partial
\rho^{2}}+\frac{5}{\rho}\frac{\partial}{\partial\rho}-\frac{K\left(
K+4\right)  }{\rho^{2}}\right]  .\label{eq:008A}%
\end{equation}
The potential energy$V_{c,\widetilde{c}}\left(  \rho,\widetilde{\rho}\right)
$ and the norm kernel $N_{c,\widetilde{c}}\left(  \rho,\widetilde{\rho
}\right)  $ can be obtained with the help of the  projection operator
$\widehat{P}_{c}$ which is presented in Ref. \cite{2018PhRvC..98b4325V}.

It is well-known that the Pauli principle leads to a nonlocal form of
potential energy operator and to appearance of the energy dependent part in
the effective potential (the right-hand side of equations (\ref{eq:008})). To
simplify solving a set of equations (\ref{eq:008}), we employ a full set of
cluster oscillator functions to expand the total wave function
\[
\Psi_{JM_{J}}^{E}=\sum_{n_{\rho},c}C_{n_{\rho},c}^{E,J}\left\vert n_{\rho
},c,J\right\rangle .
\]
This reduces a set of integro-differential equations (\ref{eq:008}) to an
algebraic form, i.e. to the system of linear algebraic equations%
\begin{equation}
\sum_{\widetilde{n}_{\rho},\widetilde{c}}\left[  \left\langle n_{\rho
},c,J\left\vert \widehat{H}\right\vert \widetilde{n}_{\rho},\widetilde
{c},J\right\rangle -E\left\langle n_{\rho},c,J|\widetilde{n}_{\rho}%
,\widetilde{c},J\right\rangle \right]  C_{\widetilde{n}_{\rho},\widetilde{c}%
}^{E,J}=0.\label{eq:009}%
\end{equation}
Cluster oscillator functions for a three-cluster configuration $A=A_{1}%
+A_{2}+A_{3}$ are determined as%
\begin{eqnarray}
&  & \left\vert n_{\rho},c,J\right\rangle =\left\vert n_{\rho},K;\lambda
,l;L;J\right\rangle \label{eq:010}\\
&  =&\widehat{\mathcal{A}}\left\{  \left[  \Phi_{1}\left(  A_{1},b_{1}%
,s_{1}\right)  \Phi_{2}\left(  A_{2},b_{2},s_{2}\right)  \Phi_{3}\left(
A_{3},b_{3},s_{3}\right)  \right]  _{S}\right.  \nonumber\\
&  \times& \left.  R_{n_{\rho}K}\left(  \rho,b\right)  \mathcal{Y}_{c}\left(
\Omega_{5}\right)  \right\}  _{JM_{J}},\nonumber
\end{eqnarray}
where $R_{n_{\rho},K}\left(  \rho,b\right)  $ is an oscillator function%
\begin{eqnarray}
  R_{n_{\rho},K}\left(  \rho,b\right)  
&  =&\left(  -1\right)  ^{n_{\rho}}\mathcal{N}_{n_{\rho},K}r^{K}\exp\left\{
-\frac{1}{2}r^{2}\right\}  L_{n_{\rho}}^{K+3}\left(  r^{2}\right)
,\label{eq:010A}\\
r &  =&\rho/b,\quad\mathcal{N}_{n_{\rho},K}=b^{-3}\sqrt{\frac{2\Gamma\left(
n_{\rho}+1\right)  }{\Gamma\left(  n_{\rho}+K+3\right)  }},\nonumber
\end{eqnarray}
$L_{n}^{\alpha}\left(  x\right)$ is the Laguerre polynomial and $b$ is an oscillator length.

System of equations (\ref{eq:009}) can be solved numerically by imposing
restrictions on the number of hyperradial excitations $n_{\rho}$ and on the
number of hyperspherical channels $c_{1}$, $c_{2}$, \ldots, $c_{N_{ch}}$. The
diagonalization procedure may be used to determine energies and wave functions
of the bound states. However, the proper boundary conditions have to be
implemented to calculate elements of the scattering $S$-matrix and
corresponding wave functions of continuous spectrum. \ Such boundary conditions and
their implementation in the proposed discreet scheme has been discussed in
Ref. \cite{2001PhRvC..63c4606V}. By solving the system of equations
(\ref{eq:009}), we obtain $N_{ch}$ wave functions $\left\{  C_{n_{\rho}%
,c}^{E,J}\right\}  $ and $N_{ch}^{2}$\ elements of the unitary $S$ matrix
$S_{c,\widetilde{c}}$. By employing the Breit-Wigner formula for a resonance
state, we deduce the energy, the total and partial widths of such state in
three-cluster continuum.

Having obtained the expansion coefficients for any state of the three-cluster
system, we can easily construct its wave function in the coordinate space. It
can be done, the first of all, for the  hyperradial wave function
\begin{equation}
\phi_{c}^{E,J}\left(  \rho\right)  =\sum_{n_{\rho}}C_{n_{\rho},c}%
^{E,J}R_{n_{\rho},K}\left(  \rho,b\right)  .\label{eq:033}%
\end{equation}
It can be also done for the total inter-cluster wave function%
\begin{equation}
\psi^{E,J}\left(  \mathbf{x},\mathbf{y}\right)  =\sum_{n_{\rho},c}C_{n_{\rho
},c}^{E,J}R_{n_{\rho},K}\left(  \rho,b\right)  \mathcal{Y}_{c}\left(
\Omega_{5}\right)  .\label{eq:034}%
\end{equation}

To get more information about the state under consideration we will study
different quantities which can be obtained with the wave function in discrete
or coordinate spaces. With wave functions in the discrete oscillator quantum
number representation we can determine a weight $W_{sh}$ of the oscillator
function belonging to the oscillator shell $N_{sh}$ in this wave function:
\begin{equation}
W_{sh}\left(  N_{sh}\right)  =\sum_{n_{\rho},c\in N_{sh}}\left\vert
C_{n_{\rho},c}^{E,J}\right\vert ^{2}, \label{eq:035}%
\end{equation}
where the summation is performed over all quantum numbers of hyperspherical harmonics and
hyperradial excitations obeying the following condition:
\[
N_{os}=2n_{\rho}+K.
\]
Here the number of oscillator quanta $N_{os}$ is fixed. 

Basis wave functions (\ref{eq:010A}) belongs to the oscillator shell with the
number of oscillator quanta $N_{os}=2n_{\rho}+K$, then it is convenient to
numerate the oscillator shells by $N_{sh}$ ( = 0, 1, 2, \ldots), which we
determine as
\[
N_{os}=2n_{\rho}+K=2N_{sh}+K_{\min},
\]
where $K_{\min}=L$ for normal parity states $\pi=\left(  -1\right)  ^{L}$ and
$K_{\min}=L+1$ for abnormal parity states $\pi=\left(  -1\right)  ^{L+1}$.
Thus we account oscillator shells starting from a "vacuum" shell ($N_{sh}$ =
0) with minimal value of the hypermomentum $K_{\min}$ compatible with a given
total orbital momentum  $L$ and the parity $\pi$.

The weights $W_{sh}$ can be calculated both for bound and resonance states.
For a bound state, the wave function $\Psi_{E,J}$ is normalized by the
condition%
\begin{equation}
\left\langle \Psi_{E,J}|\Psi_{E,J}\right\rangle =\sum_{n_{\rho},c}\left\vert
C_{n_{\rho},c}^{E,J}\right\vert ^{2}=1,\label{eq:036A}%
\end{equation}
and this quantity $W_{sh}$ determines the probability. For the continuous
spectrum state, when the wave function $\Psi_{E,J}$ is normalized by the
condition%
\begin{equation}
\left\langle \Psi_{E,J}|\Psi_{\widetilde{E},J}\right\rangle =\sum_{n_{\rho}%
,c}C_{n_{\rho},c}^{E,J}C_{n_{\rho},c}^{\widetilde{E},J}=\delta\left(
k-\widetilde{k}\right)  ,\label{eq:036B}%
\end{equation}
this quantity has a different meaning. It determines the relative contribution
of the different oscillator shells and also the shape of the resonance wave
function in the oscillator representation.

It is worthwhile to notice that oscillator functions have some important
features. For instance, oscillator functions belonging to an oscillator shell
$N_{sh}$ allow one to describe a many-particle system in a finite range of
hyperradius $0<\rho\leq b\sqrt{4N_{sh}+2K_{\min}+3}$. Outside this region,
these oscillator functions give a negligible small contribution to the 
many-particle wave function. This statement is, for example, demonstrated in
Ref. \cite{PhysRevC.97.064605}. Besides, oscillator wave functions belonging to the
oscillator shell $N_{sh}$ yield the mass root-mean-square radius which equals
$b\sqrt{\left[  2N_{sh}+K_{\min}+3\left(  A-1\right)  /2\right]  /A}$. Thus,
oscillator functions with a small value of $N_{sh}$ describe very compact
configurations of a three-cluster system with small distances between clusters.
When $N_{sh}$ is large, the oscillator functions represent dispersed
three-cluster configurations. There are two principal regimes in these
configurations. The first regime is associated with a two-body type of
asymptotic when two clusters are at a small distance and the third cluster is
moved far away. The second regime accounts for the case when all three
clusters are well separated. Taking these into account, we will deduce from
analysis of shell weights $W_{sh}$ whether a wave function of a bound or
resonance state describes a compact or dispersed three-cluster configuration.

The weights oscillator shells $W_{sh}$ have been used by different authors 
to study wave functions obtained in different microscopic models. For example, 
they where employed in Refs. \cite{2008IJMPE..17.2005N,2009FBS....45..145N} to 
study the Hoyle states in $^{12}$C within  the fermionic molecular dynamics. 
Note that our definition of the weights $W_{sh}$ is consistent with more general 
definition suggested by Y. Suzuki et al in Ref. \cite{1996PhRvC..54.2073S}.

Important ingredients of the present microscopic model are a nucleon-nucleon potential and the Coulomb interaction. As we pointed out above, we formulated our model in such a way that the cluster-cluster interaction originated from the nucleon-nucleon potential are the same in mirror nuclei. Thus difference in the position and width of resonance states  is due to the Coulomb interaction.  Let us consider in more detail the Coulomb potential energy in mirror nuclei. For the $L$ and $R$  nuclei of $Z_{L} < Z_{R}$ it has the
form%
\begin{equation}
V_{C}^{\left(  L\right)  }=\sum_{i<j}^{Z_{L}}\frac{e^{2}}{\left\vert
\mathbf{r}_{i}-\mathbf{r}_{j}\right\vert },\quad V_{C}^{\left(  R\right)
}=\sum_{i<j}^{Z_{R}}\frac{e^{2}}{\left\vert \mathbf{r}_{i}-\mathbf{r}%
_{j}\right\vert }. \label{eq:C20}%
\end{equation}
It is easy to see that the potential energy for the $R$ nucleus, where
$Z_{R}>Z_{L}$, can be represented as a combination of three components%
\begin{eqnarray}
V_{C}^{\left(  R\right)  } &  =&\sum_{i<j\in Z_{R}}\frac{e^{2}}{\left\vert
\mathbf{r}_{i}-\mathbf{r}_{j}\right\vert }=\sum_{i<j\in Z_{L}}\frac{e^{2}%
}{\left\vert \mathbf{r}_{i}-\mathbf{r}_{j}\right\vert }\label{eq:C21}\\
&  +& \sum_{i\in Z_{L}}\sum_{j\in\left(  Z_{R}-Z_{L}\right)  }\frac{e^{2}%
}{\left\vert \mathbf{r}_{i}-\mathbf{r}_{j}\right\vert }+\sum_{j>i\in\left(
Z_{L}-Z_{R}\right)  }\frac{e^{2}}{\left\vert \mathbf{r}_{i}-\mathbf{r}%
_{j}\right\vert }. \nonumber
\end{eqnarray}
In the right hand side of equation (\ref{eq:C21}), the first component is 
the Coulomb potential energy of the $L$ nucleus, the
third component is the potential energy of extra protons (with respect to
protons of the $L$ nucleus), and the second component represents the potential
energy of the interaction of extra protons with protons of the $L$ nucleus.
The last two components, which we denote as
\begin{eqnarray}
\Delta V_{C}  &  =&V_{C}^{\left(  R\right)  }-V_{C}^{\left(  L\right)
}\label{eq:C21D}\\
&  =&\sum_{i\in Z_{L}}\sum_{j\in\left(  Z_{R}-Z_{L}\right)  }\frac{e^{2}%
}{\left\vert \mathbf{r}_{i}-\mathbf{r}_{j}\right\vert }+\sum_{j>i\in\left(
Z_{L}-Z_{R}\right)  }\frac{e^{2}}{\left\vert \mathbf{r}_{i}-\mathbf{r}%
_{j}\right\vert },\nonumber
\end{eqnarray}
determine the shift of bound and resonance states in the $R$ nucleus with
respect to their position in the $L$ nucleus. In all but one nuclei we have
only one extra proton, thus for a pair of nuclei $^{7}$Li and $^{7}$Be, $^{9}%
$Be and $^{9}$B, $^{11}$B and $^{11}$C only the second component in Eq.
(\ref{eq:C21}) determines the Coulomb shift. While for nuclei $^{8}$Li and
$^{8}$B, where there are two extra protons, the second and the third
components take part in shifting of parameters of bound and resonance states.

Taking into account a three-cluster structure of mirror nuclei, we obtain an
alternative way to present the differences of the Coulomb interaction. The
Coulomb potential energy for both the  $L$ and $R$ nuclei can be represented as%
\begin{eqnarray}
V_{C}  &  =& \sum_{c=1}^{3}\sum_{i<j\in Z_{c}}\frac{e^{2}}{\left\vert
\mathbf{r}_{i}-\mathbf{r}_{j}\right\vert }+\sum_{i\in Z_{1}}\sum_{j\in Z_{2}%
}\frac{e^{2}}{\left\vert \mathbf{r}_{i}-\mathbf{r}_{j}\right\vert
}\label{eq:C22}\\
&  +& \sum_{i\in Z_{2}}\sum_{j\in Z_{3}}\frac{e^{2}}{\left\vert \mathbf{r}%
_{i}-\mathbf{r}_{j}\right\vert }+\sum_{i\in Z_{1}}\sum_{j\in Z_{3}}\frac
{e^{2}}{\left\vert \mathbf{r}_{i}-\mathbf{r}_{j}\right\vert }\nonumber
\end{eqnarray}
The first three terms of eq. (\ref{eq:C22}) represent the internal Coulomb
potential energy of a cluster $c$ ($c$=1, 2, 3) and the last three-terms are
the Coulomb interactions of different clusters. It is obvious, that the
internal Coulomb energy is nonzero for a cluster containing 2 and more
protons, and the Coulomb interactions between different clusters \ are nonzero
when both interacting clusters contain one and more protons. Note that in such
representation (\ref{eq:C22}) of the Coulomb potential energy, the Coulomb
potential energy difference $\Delta V_{C}$ may originates from the difference
of the internal energy of a cluster and from the interaction between clusters.
It can be represented as%
\begin{eqnarray}
\Delta V_{C} &  =& \sum_{c=1}^{3}\sum_{i<j\in Z_{c,R}-Z_{c,L}}\frac{e^{2}%
}{\left\vert \mathbf{r}_{i}-\mathbf{r}_{j}\right\vert } \nonumber \\
&+&\sum_{i\in
Z_{1,R}-Z_{1,L}}\sum_{j\in Z_{2,R}-Z_{2,L}}\frac{e^{2}}{\left\vert
\mathbf{r}_{i}-\mathbf{r}_{j}\right\vert }\label{eq:C22D}\\
&  +& \sum_{i\in Z_{2,R}-Z_{2,L}}\sum_{j\in Z_{3,R}-Z_{3,L}}\frac{e^{2}%
}{\left\vert \mathbf{r}_{i}-\mathbf{r}_{j}\right\vert } \nonumber \\
&+&\sum_{i\in
Z_{1,R}-Z_{1,L}}\sum_{j\in Z_{3,R}-Z_{3,L}}\frac{e^{2}}{\left\vert
\mathbf{r}_{i}-\mathbf{r}_{j}\right\vert }.\nonumber
\end{eqnarray}

Let us consider the nuclei in our hands. The nuclei $^{7}$Li and $^{7}$Be, as
was shown in Table \ref{Tab:MirrorNuclei}, differ in one proton. Thus, in
$^{7}$Be we have additional terms caused by the Coulomb interaction, namely,
the interaction of a valence proton with alpha-particle and the interaction of
that proton with a deuteron. The similar picture is observed in a pair $^{9}%
$Be and $^{9}$B, where $\Delta V_{C}$ consists of the interaction of the
valence proton with the first and second alpha-particles. In nuclei $^{8}$Li
and $^{8}$B and $^{11}$B and $^{11}$C we have got a rather
different situation. In these nuclei, the Coulomb interaction contribute to
the internal energy of the cluster $^{3}$He and makes a stronger interaction
(repulsion) between $^{3}$He and alpha-particle(s) in $^{8}$B ($^{11}$C) with
respect to the $t-\alpha$ interaction in $^{8}$Li ($^{11}$B).

The Coulomb potential energy difference $\Delta V_{C}$ can be treated as a
perturbation and thus the Coulomb shift can be evaluated by using wave
functions of the $L$ nucleus, or, by introducing a factor $\lambda_{C}$ and
considering the interaction $\lambda_{C}\Delta V_{C}$, one can study the
trajectory of bound and resonance states when the parameter $\lambda_{C}$ is
changed from zero to one. However, in our calculations the energies and widths
and wave functions of resonance states in both $L$ and $R$ nuclei are obtained
in the same way with the corresponding boundary conditions.

In the next Section we will study how the Coulomb potential energy difference
$\Delta V_{C}$ changes the energy and width of resonance states in the $R$
nucleus with respect to its position in the $L$ nucleus.

\section{Effects of Coulomb forces}

\label{Sec:Effects}

We will not discuss details of calculations as they were discussed in papers
mentioned in Table \ref{Tab:MirrorNuclei}. We just outline some general steps
of these calculations. In our calculations we use a common oscillator length
$b$ for all interacting clusters. The oscillator length was chosen to minimize
the energy of the three-cluster threshold. This optimizes a description of the
internal structure of clusters. The Minnesota potential (MP) \cite{kn:Minn_pot1} and the modified
Hasegawa-Nagata potential (MHNP) \cite{potMHN1, potMHN2} were involved in all calculations. The
exchange parameter $u$ of the MP and the Majorana parameter $m$ of the MHNP
was slightly adjusted to reproduce the ground state energy of the $L$ nucleus.
The same values of $m$ or $u$ were used for the $R$ nucleus. In this case, the
interactions between clusters, generated by the nucleon-nucleon interaction,
are the same in the $L$ and $R$ nuclei.

In this paper we do not compare our results with the available experimental
data, as it was done in the references mentioned in Table
\ref{Tab:MirrorNuclei}. However, we will compare our results with the results
of other theoretical approaches.

\subsection{$^{9}$Be and $^{9}$B}

Spectrum of resonance states in $^{9}$Be and $^{9}$B has been obtained in
Refs. \cite{2017PhRvC..96c4322V, 2014PAN..77.555N} within the AMHHB.
This method was selected to study parameters and nature of resonance states in
$^{9}$Be and $^{9}$B because all resonance states of these nuclei are embedded
in three-cluster continuum and this method implements proper boundary
conditions for the three-cluster continuous spectrum states.

In Table \ref{Tab:Resons9Be9BMHNP} we demonstrate energies and widths of
resonance states in $^{9}$Be and $^{9}$B. They were obtained in Ref.
\cite{2017PhRvC..96c4322V} with the MHNP within the AMHHB. Table
\ref{Tab:Resons9Be9BMHNP} also displays the Coulomb shift $R_{C}$ and
rotational angle $\theta_{C}$.%

\begin{table}[htbp] \centering
\caption{Spectrum of bound and resonance states in $^9$Be and $^9$B calculated with
the MHNP. Energies $E$ and widths $\Gamma$ are in MeV.}%
\begin{tabular}
[c]{ccccccc}\hline
& \multicolumn{2}{c}{$^{9}$Be} & \multicolumn{2}{c}{$^{9}$B} &
\multicolumn{2}{c}{Coulomb shift}\\\hline
$J^{\pi}$ & $E$ & $\Gamma$ & $E$ & $\Gamma$ & $R_{C}$ & $\theta_{C}%
$\\\hline
$3/2_{1}^{-}$ & -1.574 & 0.00 & 0.379 & 1.1$\times$10$^{-6}$ & 1.953 &
3.23$\times$10$^{-5}$\\
$1/2_{1}^{+}$ & 0.338 & 0.168 & 0.636 & 0.477 & 0.429 & 46.04\\
$5/2_{1}^{-}$ & 0.897 & 2.4$\times$10$^{-5}$ & 2.805 & 0.018 & 1.908 & 0.54\\
$1/2_{1}^{-}$ & 2.866 & 1.597 & 3.398 & 3.428 & 1.907 & 73.80\\
$5/2_{1}^{+}$ & 2.086 & 0.112 & 3.670 & 0.415 & 1.613 & 10.83\\
$3/2_{1}^{+}$ & 4.062 & 1.224 & 4.367 & 3.876 & 2.669 & 83.44\\
$3/2_{2}^{-}$ & 2.704 & 2.534 & 3.420 & 3.361 & 1.094 & 49.12\\
$7/2_{1}^{-}$ & 4.766 & 0.404 & 6.779 & 0.896 & 2.072 & 13.74\\
$9/2_{1}^{+}$ & 4.913 & 1.272 & 6.503 & 2.012 & 1.754 & 24.96\\
$5/2_{2}^{-}$ & 5.365 & 4.384 & 5.697 & 5.146 & 0.831 & 66.46\\
$7/2_{1}^{+}$ & 5.791 & 3.479 & 7.100 & 4.462 & 1.637 & 36.90\\\hline
\end{tabular}
\label{Tab:Resons9Be9BMHNP}%
\end{table}%

Let us look closely what possible scenarios are realized in nuclei $^{9}$Be
and $^{9}$B and how it depends on the total angular momentum $J$.%

\begin{figure}
[ptb]
\begin{center}
\includegraphics[width=\columnwidth]
{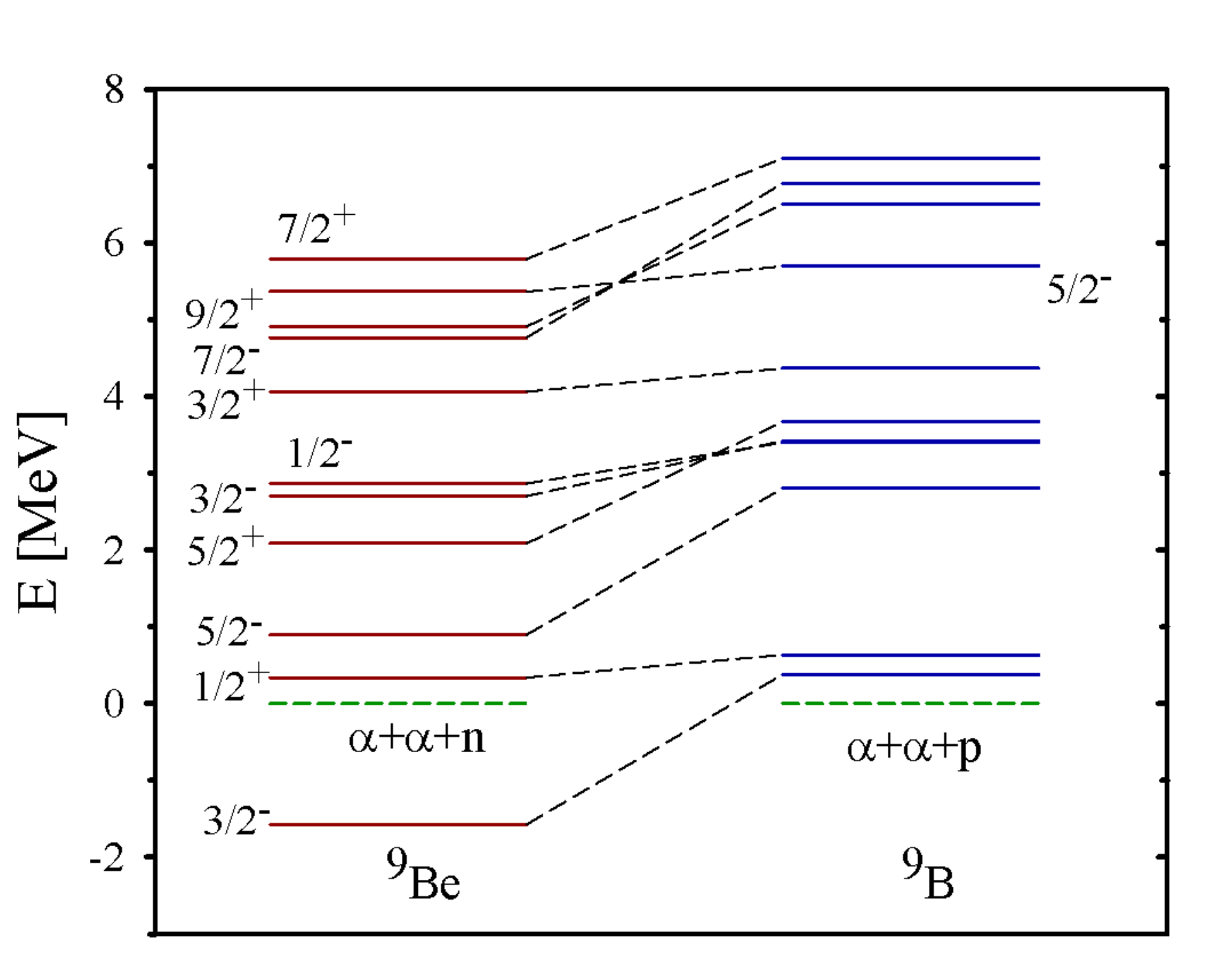}%
\caption{Spectra of bound and resonance states in $^{9}$Be and $^{9}$B.}%
\label{Fig:Spectr9Be&9BMHNP}%
\end{center}
\end{figure}

The first effects of the Coulomb forces in the mirror nuclei $^{9}$Be and $^{9}%
$B \ can be seen in Fig. \ref{Fig:Spectr9Be&9BMHNP} where spectra of these
nuclei are shown. Five dashed lines, connecting levels with the same total
angular momentum $J$ and parity $\pi$ in $^{9}$Be and $^{9}$B, show that
Coulomb forces significantly shift up levels ($J^{\pi}$= 3/2$^{-}$, 5/2$^{-}$,
5/2$^{+}$, 7/2$^{-}$ and 9/2$^{-}$) and four dashed lines indicate a moderate
shift up of energy of resonance states ($J^{\pi}$= 1/2$^{+}$, 3/2$^{-}$,
1/2$^{-}$ and 3/2$^{+}$) in $^{9}$B comparing with correspondent states in
$^{9}$Be.

To see effects of the Coulomb interaction more vividly we present Fig.
\ref{Fig:DEvsDG9Be9B}.  In this figure and other figures below, the arcs (grey
dashed curves) mark the Coulomb shift $R_{C}$=1, 2 and 3 MeV, and a set of
rays (grey solid lines) indicate the Coulomb rotational angles $\theta_{C}$
every 15 degrees. As we can see the largest group of resonance states are
concentrated around $R_{C}$= 2 and almost all \ states of this group except
for one lie below $\theta_{C}$=45$^{\circ}$.%

\begin{figure}[ptbh]
\begin{center}
\includegraphics[width=\columnwidth]
{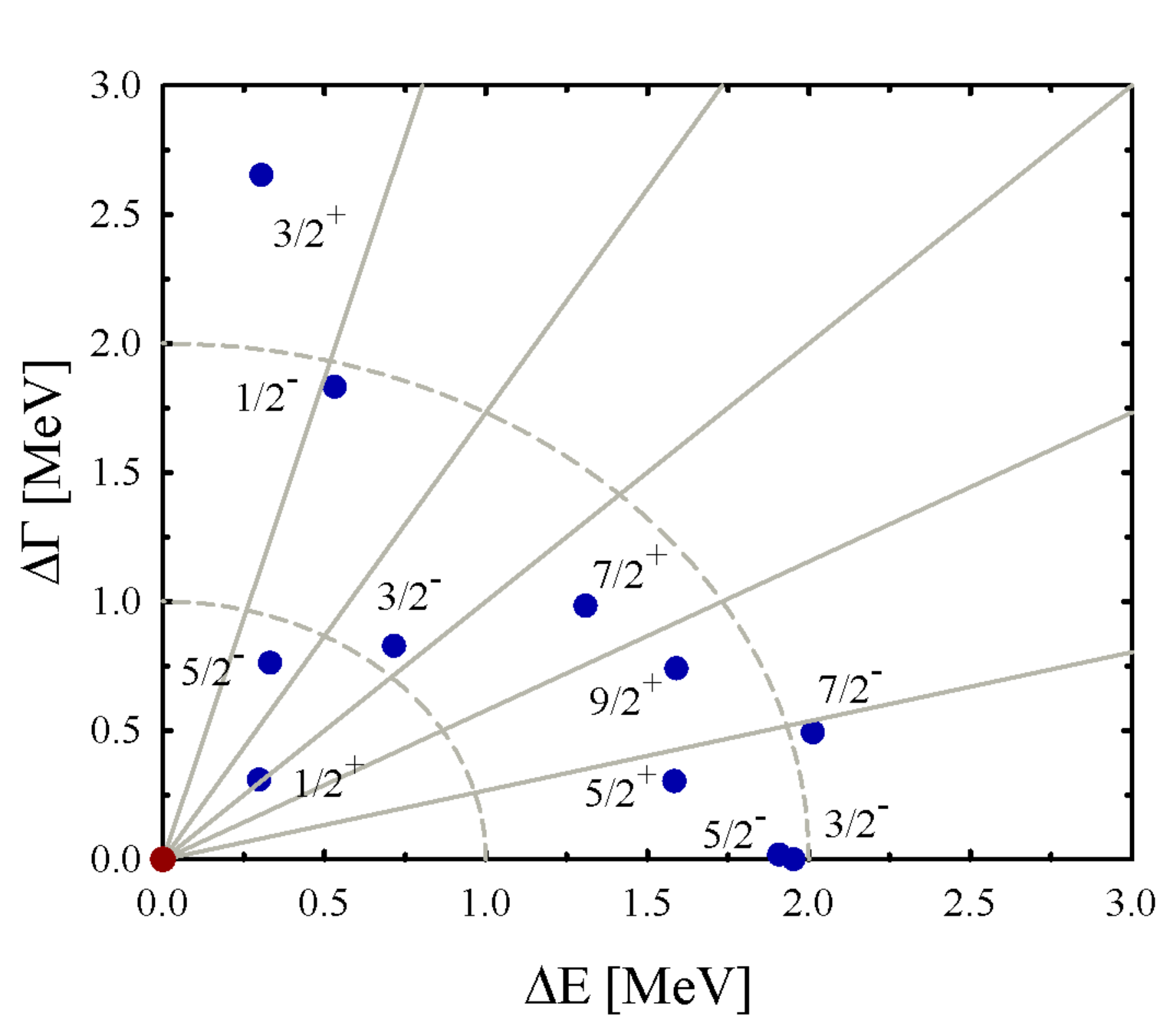}%
\caption{The shift and rotation of resonance states in $^{9}$B caused by the
Coulomb interactions. Counterparts of these resonance states in $^{9}$Be are
put in the origin of coordinates.}%
\label{Fig:DEvsDG9Be9B}%
\end{center}
\end{figure}

Thus, the Coulomb interaction has week (the first group of 
$\Delta E=0.25\sim 0.75$ MeV), moderate (the second group of 
$\Delta E=1.25\sim 1.8$ MeV) or strong (the third group of $\Delta E > 1.9$ MeV) influence on
parameters of resonance states in mirror nuclei $^{9}$Be and $^{9}$B. And these three groups 
are observed in terms of the Coulomb shift $R_{C}$ and the Coulomb angle
$\theta_{C}$. We also observed that the first scenario is realized in these
nuclei as the Coulomb interaction increases both energy and width of resonance
states in $^{9}$B with respect to their values in $^{9}$Be.

\subsubsection{CSM}

\begin{figure}[ptb]
\begin{center}
\includegraphics[width=\columnwidth]
{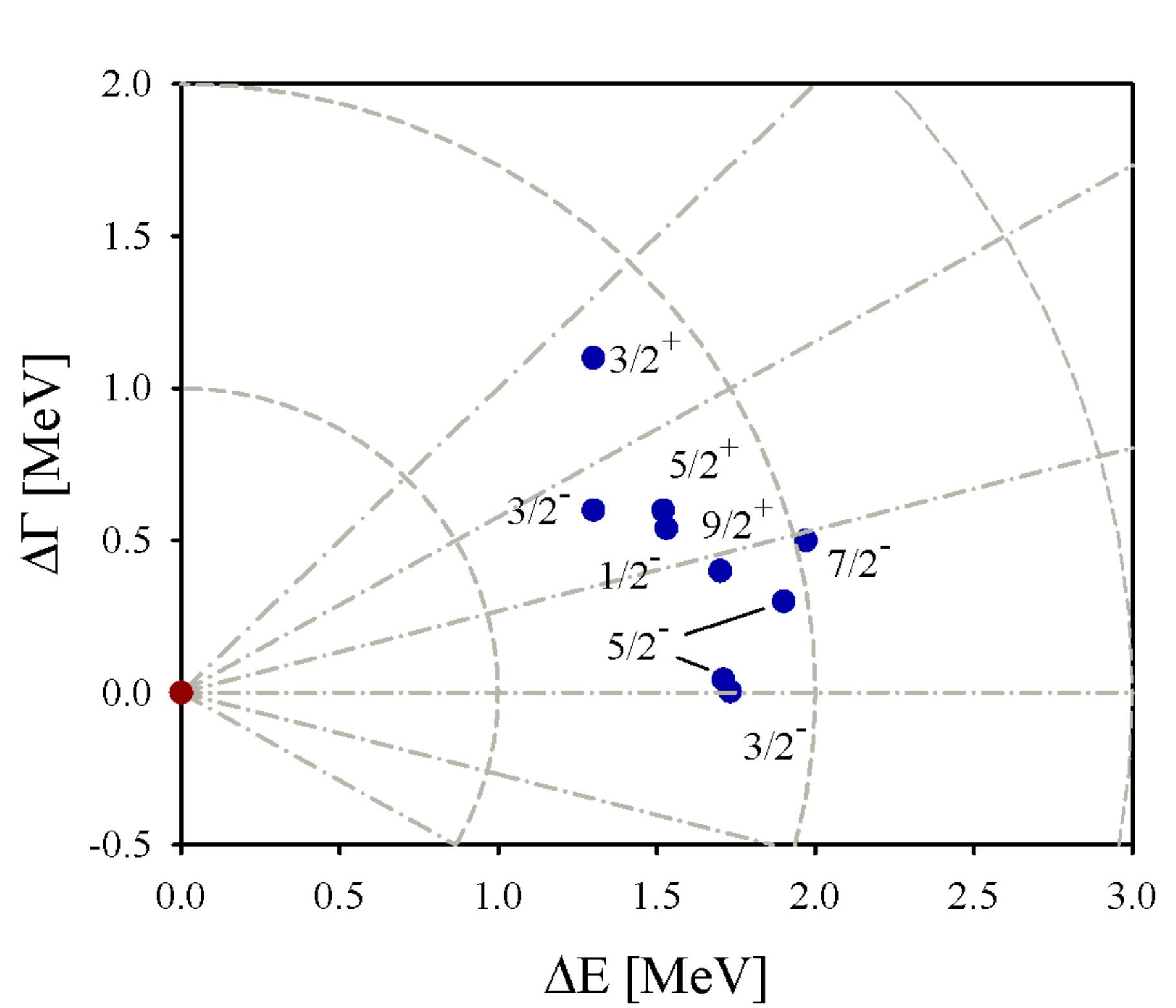}%
\caption{The Coulomb shift of resonance states in $^{9}$B with respect to
those in $^{9}$Be, obtained with the complex scaling method. The data are 
taken from Ref. \cite{1996PhRvC..54..132A}.}%
\label{Fig:DEvsDG9Be9BArai}%
\end{center}
\end{figure}

The complex scaling method (CSM) has been used in Ref.
\cite{1996PhRvC..54..132A} to determine energies and widths of resonance
states in mirror nuclei $^{9}$Be and $^{9}$B. Parameters of resonance states
were obtain with the MP. The detail comparison of results of the CSM with the
AMHHB was carried out in Ref. \cite{2014PAN..77.555N}. Here we wish to present
the results of CSM on the $E-\Gamma$ plain in order to see explicitly effects
of the Coulomb interactions detected within this method. We display these
results in Fig. \ref{Fig:DEvsDG9Be9BArai}. As we see, all resonance states lie
between $R_{C}$=1.25 and $R_{C}$=2.0 MeV and this results is consistent with
results of the AMHHB displayed in Fig. \ref{Fig:DEvsDG9Be9B}. However,
contrary to the first group of the results in the AMHHB, there are no resonance states with the weak effects of
the Coulomb interactions \ ($R_{C} \approx$%
1 MeV) in the CSM. The Coulomb shift angles $\theta_{C}$ in the CSM do not exceed
45$^{\circ}$ which is smaller than the values of $\theta_{C}$ in the AMHHB.
The difference between results of the CSM and the AMHHB may be ascribed to the
different methods of location of resonance states and partially to different
nucleon-nucleon potentials used in each approach.%

\subsection{$^{11}$B and $^{11}$C}

In this section we consider bound and resonance states in $^{11}$B and $^{11}%
$C. Parameters of these states were determined in Ref.
\cite{2013UkrJPh.58.544V} within the AMHHB by employing the MHNP. Some
additional information on the structure of wave functions of resonance states
in $^{11}$B and $^{11}$C can be found in Ref. \cite{2018PhRvC..98b4325V}.

Some remarks should be made about an application of the AMHHB method to the
nuclei $^{11}$B and $^{11}$C. As it was pointed out above, this method was
designed and applied to study light nuclei with a prominent three-cluster
structure. These nuclei can be easily split on three clusters and thus the
three-cluster threshold has a lowest energy among other two- and
three-cluster decay channels. However, it is not the case for nuclei $^{11}$%
B and $^{11}$C. For these nuclei,  the lowest decay channel is the binary
channel $^{7}$Li+$\alpha $ and $^{7}$Be+$\alpha $, respectively. At the
present time and with the present version of the AMHHB model, it is
difficult to incorporate simultaneously both two- and three-body channels.
Such a work is in the progress. The first step in this direction was made in
Ref. \cite{PhysRevC.97.064605} where the ability of the oscillator
three-cluster functions (\ref{eq:010}) to describe binary and ternary channels has
been investigated. In the present paper, as in Refs. \cite%
{2013UkrJPh.58.544V, 2018PhRvC..98b4325V} we disregard the binary
channels in $^{11}$B and $^{11}$C and consider the three-cluster continuum
only. In Refs. \cite{2013UkrJPh.58.544V, 2018PhRvC..98b4325V}, such
an approximation was aimed to study resonance states in the cluster
continuum of $^{11}$B and $^{11}$C and to search for the Hoyle-analog
states. It was assumed that a binary channel does not change dramatically
the spectrum and structure of three-cluster resonance states. Here we use
the same approximation and the same motivation to study effects of the
Coulomb interaction on parameters of three-cluster resonance states in the
mirror nuclei $^{11}$B and $^{11}$C.

The richest spectra of resonance states were obtained for $^{11}$B and $^{11}%
$C. We detected 20 resonance states in $^{11}$B and 18 resonance states in
$^{11}$C, at least three resonance states for a fixed values of the total
angular momentum $J$ and parity $\pi$. Such a large number of resonance states
is stipulated by huge centrifugal and Coulomb barriers.

However, we start with the spectra of bound states in $^{11}$B and $^{11}$C,
which are also very rich. Energies of bound states are shown in Table
\ref{Tab:Spect11Bvs11CBS}. As we can see, the nucleus $^{11}$B has ten bound
states, while the nucleus $^{11}$C has only seven bound states. Thus the
Coulomb interaction moves three bound states to the continuous spectrum and
transforms them into resonance states. The Coulomb shift for bound states
starts from 1.98 MeV for the ground states and slowly decreases to 1.45 MeV.
As one could expect the Coulomb shift is reduced for weakly bound states,
which have a dispersed three-cluster configuration.%

\begin{table}[tbp] \centering
\caption{Spectrum of bound states in $^{11}$B and $^{11}$C.}%
\begin{tabular}
[c]{cccc}\hline
& $^{11}$B & $^{11}$C & Coulomb shift\\\hline
$J^{\pi}$ & $E$ (MeV) & $E$ (MeV) & $R_{C}$(MeV)\\\hline
$3/2_{1}^{-}$ & -11.0520 & -9.0710 & 1.981\\
$1/2_{1}^{-}$ & -9.6440 & -7.7210 & 1.923\\
$5/2_{1}^{-}$ & -7.3770 & -5.4420 & 1.935\\
$3/2_{2}^{-}$ & -5.6630 & -3.8320 & 1.831\\
$1/2_{1}^{+}$ & -2.7590 & -1.2700 & 1.489\\
$5/2_{1}^{+}$ & -2.7360 & -1.1680 & 1.568\\
$3/2_{1}^{+}$ & -1.5320 & -0.0850 & 1.447\\
$5/2_{2}^{-}$ & -1.0173 &  & \\
$1/2_{2}^{-}$ & -0.1895 &  & \\
$5/2_{2}^{+}$ & -0.0414 &  & \\\hline
\end{tabular}
\label{Tab:Spect11Bvs11CBS}%
\end{table}%

\begin{figure}[ptbh]
\begin{center}
\includegraphics[width=\columnwidth]
{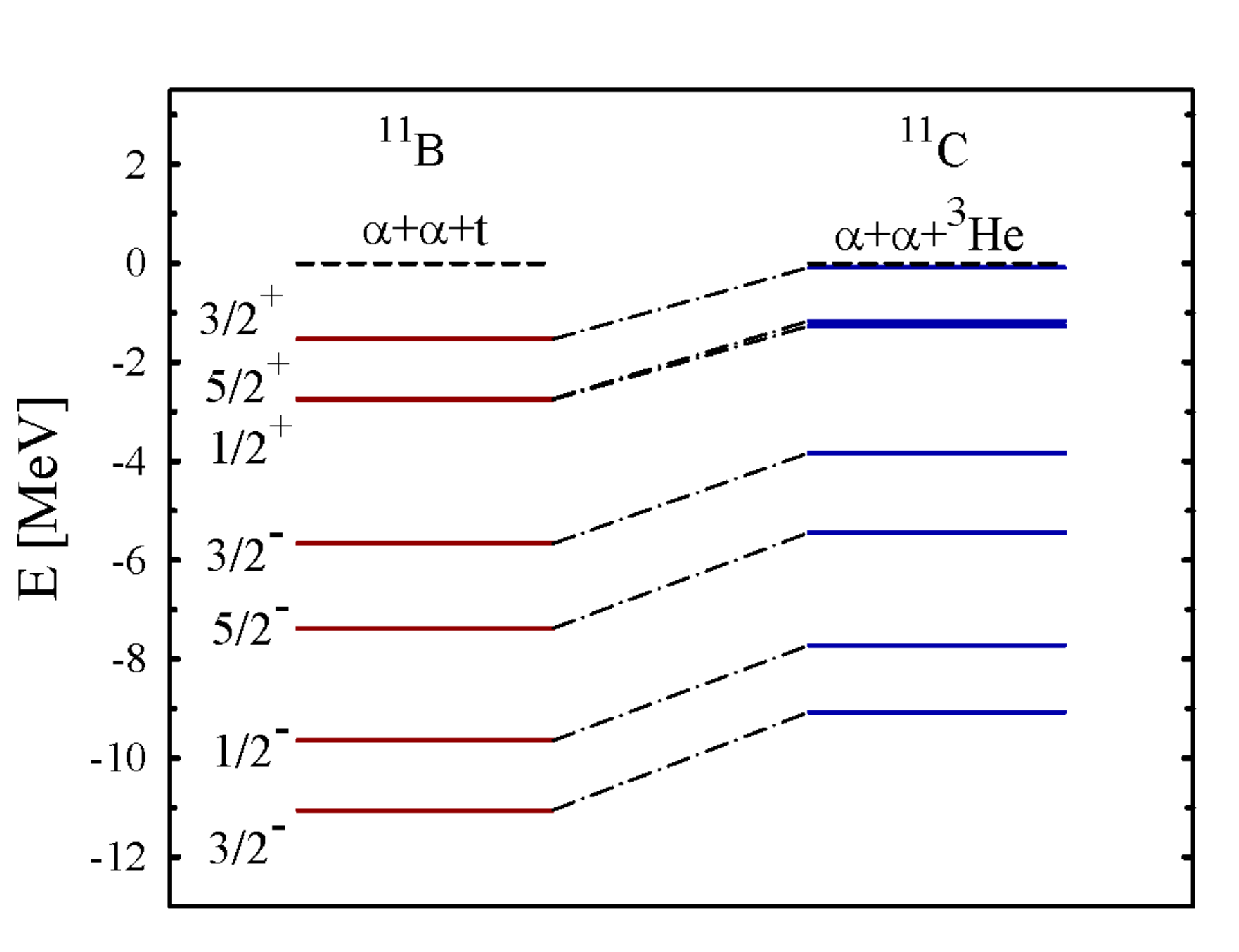}%
\caption{Spectra of bound states in $^{11}$B and $^{11}$C.}%
\label{Fig:BoundSts11B11C}%
\end{center}
\end{figure}

In Fig. \ref{Fig:BoundSts11B11C} we display how the Coulomb interaction shifts
the bound state of $^{11}$C with respect to those in $^{11}$B. This figure
demonstrates the Coulomb interaction shifts bound states of $^{11}$C
approximately on the almost same values (1.4$\sim$2.0 MeV) for all bound states.%

In Table \ref{Tab:Spect11Bvs11C} we display parameters of negative parity
resonance states in the three-cluster continuum of $^{11}$B and $^{11}$C. In
Table \ref{Tab:Spect11Bvs11C} we also included two bound states in $^{11}$B
which are transformed into resonance states in $^{11}$C due to the repulsive
Coulomb interaction.  In Table \ref{Tab:Spect11Bvs11C} we also show the
rotational $\theta_{C}$ and shift $R_{C}$ parameters caused by the Coulomb interaction.

\begin{table}[tbp] \centering
\caption{Spectrum of  the negative parity resonance states in $^{11}$B and $^{11}$C and the
Coulomb shift $R_{C}$ and rotation $\theta_{C}$ parameters. Energies $E$ are in MeV and widths $\Gamma$ are in keV.}%
\begin{tabular}
[c]{ccccccc}\hline
& \multicolumn{2}{c}{$^{11}$B} & \multicolumn{2}{c}{$^{11}$C} &
\multicolumn{2}{c}{Coulomb shifts}\\\hline
$J^{\pi}$ & $E$ & $\Gamma$ & $E$ & $\Gamma$ & $R_{C}$ & $\theta_{C}$\\\hline
3/2$_{1}^{-}$ & 0.755 & 5.8$\times$10$^{-4}$ & 0.805 & 9.9$\times$10$^{-6}$ &
0.050 & -0.65\\
3/2$_{2}^{-}$ & 1.402 & 0.185 & 1.920 & 0.105 & 0.524 & -8.78\\
3/2$_{3}^{-}$ & 1.756 & 0.143 & 2.324 & 0.619 & 0.741 & 39.96\\
1/2$_{3}^{-}$ & -0.190 & 0.0 & 1.142 & 7.1$\times$10$^{-4}$ & 1.332 & 0.03\\
1/2$_{4}^{-}$ & 1.436 & 0.374 & 2.266 & 0.790 & 0.928 & 26.62\\
1/2$_{5}^{-}$ & 1.895 & 0.101 & 3.014 & 0.366 & 1.150 & 13.32\\
1/2$_{6}^{-}$ & 2.404 & 0.450 & 3.326 & 0.383 & 0.925 & -4.18\\
5/2$_{3}^{-}$ & -1.017 & 0.0 & 0.783 & 9.6$\times$10$^{-8}$ & 1.800 &
3.05$\times$10$^{-6}$\\
5/2$_{4}^{-}$ & 0.583 & 5.1$\times$10$^{-7}$ & 1.897 & 0.006 & 1.314 & 0.26\\
5/2$_{5}^{-}$ & 1.990 & 0.032 & 3.026 & 0.183 & 1.047 & 8.29\\
5/2$_{6}^{-}$ & 2.251 & 0.138 & 3.491 & 0.393 & 1.266 & 11.62\\
7/2$_{1}^{-}$ & 1.591 & 0.004 & 2.700 & 0.067 & 1.111 & 3.25\\
7/2$_{2}^{-}$ & 1.778 & 0.003 & 3.538 & 0.021 & 1.760 & 0.59\\\hline
\end{tabular}
\label{Tab:Spect11Bvs11C}%
\end{table}%

Resonance states of the positive parity are displayed in Table
\ref{Tab:Spect11Bvs11CP}. This table contains also one bound 5/2$_{1}^{+}$
state in $^{11}$B, which is transformed into the narrow resonance state due to
the Coulomb interaction.%

\begin{table}[tbp] \centering
\caption{Energies and widths of positive parity resonance states in $^{11}$B and $^{11}$C, and the Coulomb shift parameters as well. Energies $E$ are in MeV and widths $\Gamma$ are in keV.}%
\begin{tabular}
[c]{ccccccc}\hline
& \multicolumn{2}{c}{$^{11}$B} & \multicolumn{2}{c}{$^{11}$C} &
\multicolumn{2}{c}{Coulomb shifts}\\\hline
$J^{\pi}$ & $E$ & $\Gamma$ & $E$ & $\Gamma$ & $R_{C}$ & $\theta_{C}$\\\hline
1/2$_{1}^{+}$ & 0.437 & 0.015 & 0.906 & 0.162 & 0.492 & 17.40\\
1/2$_{2}^{+}$ & 0.702 & 0.012 & 1.930 & 0.059 & 1.229 & 2.19\\
1/2$_{3}^{+}$ & 1.597 & 0.016 & 2.679 & 0.086 & 1.084 & 3.70\\
3/2$_{1}^{+}$ & 1.147 & 1.49$\times$10$^{-3}$ & 2.268 & 0.034 & 1.121 &
1.671\\
3/2$_{2}^{+}$ & 1.367 & 8.58$\times$10$^{-3}$ & 2.478 & 0.159 & 1.121 & 7.73\\
3/2$_{3}^{+}$ & 1.715 & 4.12$\times$10$^{-2}$ & 2.850 & 0.115 & 1.137 & 3.73\\
5/2$_{1}^{+}$ & -0.041 & 0.0 & 1.460 & 9.00$\times$10$^{-4}$ & 1.502 & 0.03\\
5/2$_{2}^{+}$ & 1.047 & 1.54$\times$10$^{-3}$ & 2.346 & 8.27$\times$10$^{-2}$
& 1.302 & 3.58\\
5/2$_{3}^{+}$ & 1.951 & 4.02$\times$10$^{-2}$ & 3.179 & 0.123 & 1.231 &
3.85\\\hline
\end{tabular}
\label{Tab:Spect11Bvs11CP}%
\end{table}%

\begin{figure}[ptb]
\begin{center}
\includegraphics[width=\columnwidth]
{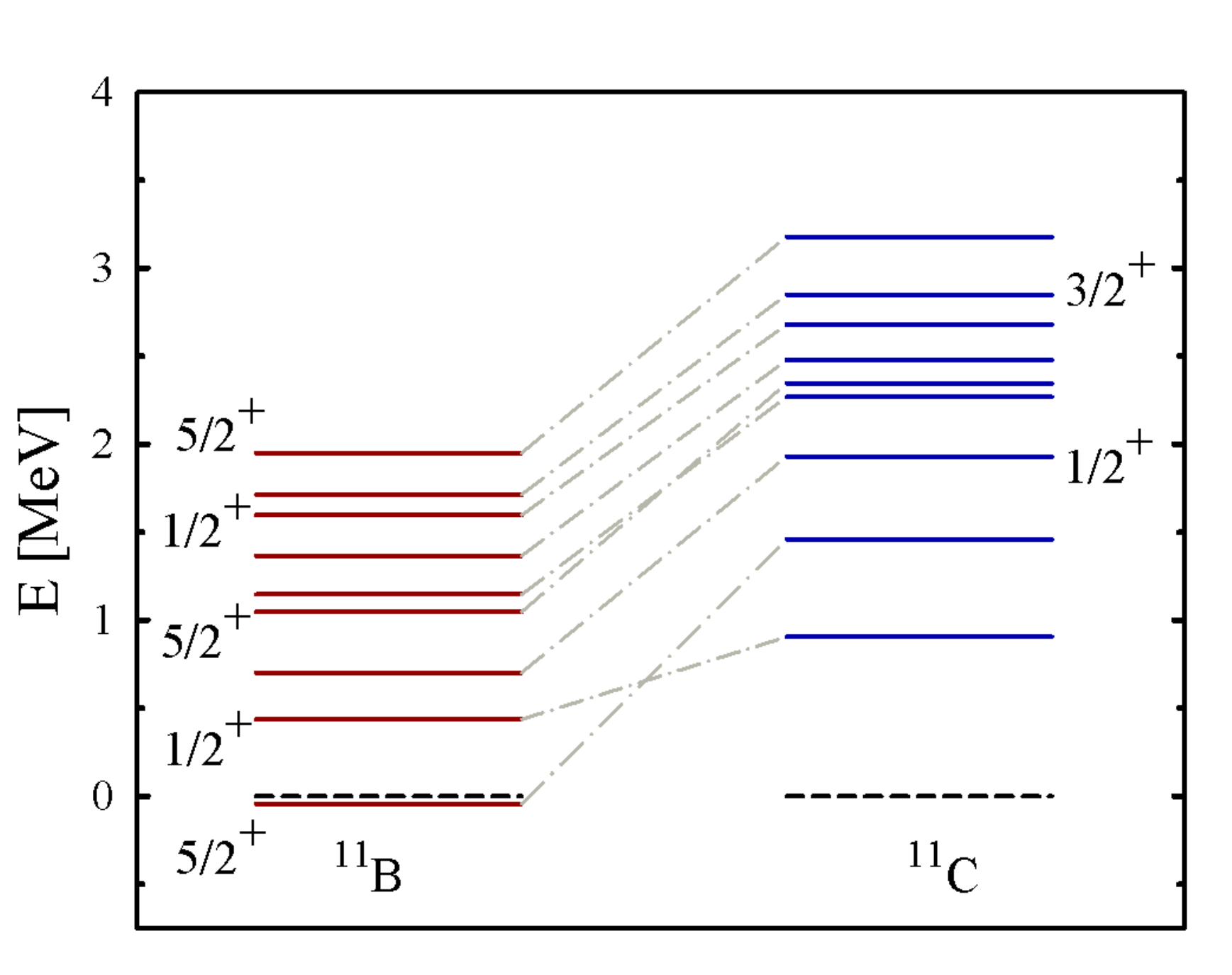}%
\caption{Spectra of the positive parity resonance states in $^{11}$B and
$^{11}$C.}%
\label{Fig:Spect11B11CRSP}%
\end{center}
\end{figure}

\begin{figure}[ptb]
\begin{center}
\includegraphics[width=\columnwidth]
{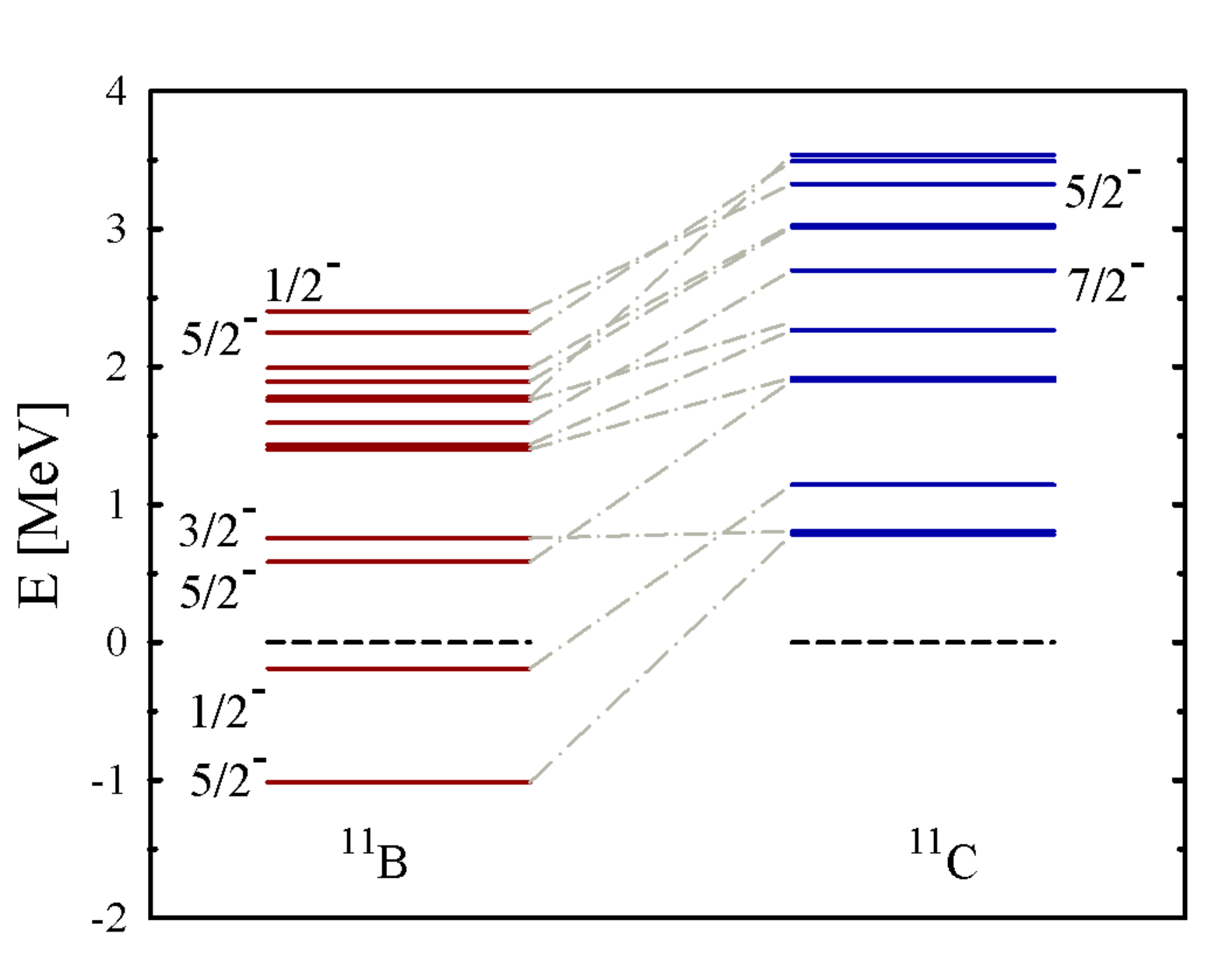}%
\caption{Spectra of negative parity resonance states in the mirror nuclei $^{11}$B and
$^{11}$C.}%
\label{Fig:Spect11B11CRSN}%
\end{center}
\end{figure}

\begin{figure}[ptbh]
\begin{center}
\includegraphics[width=\columnwidth]
{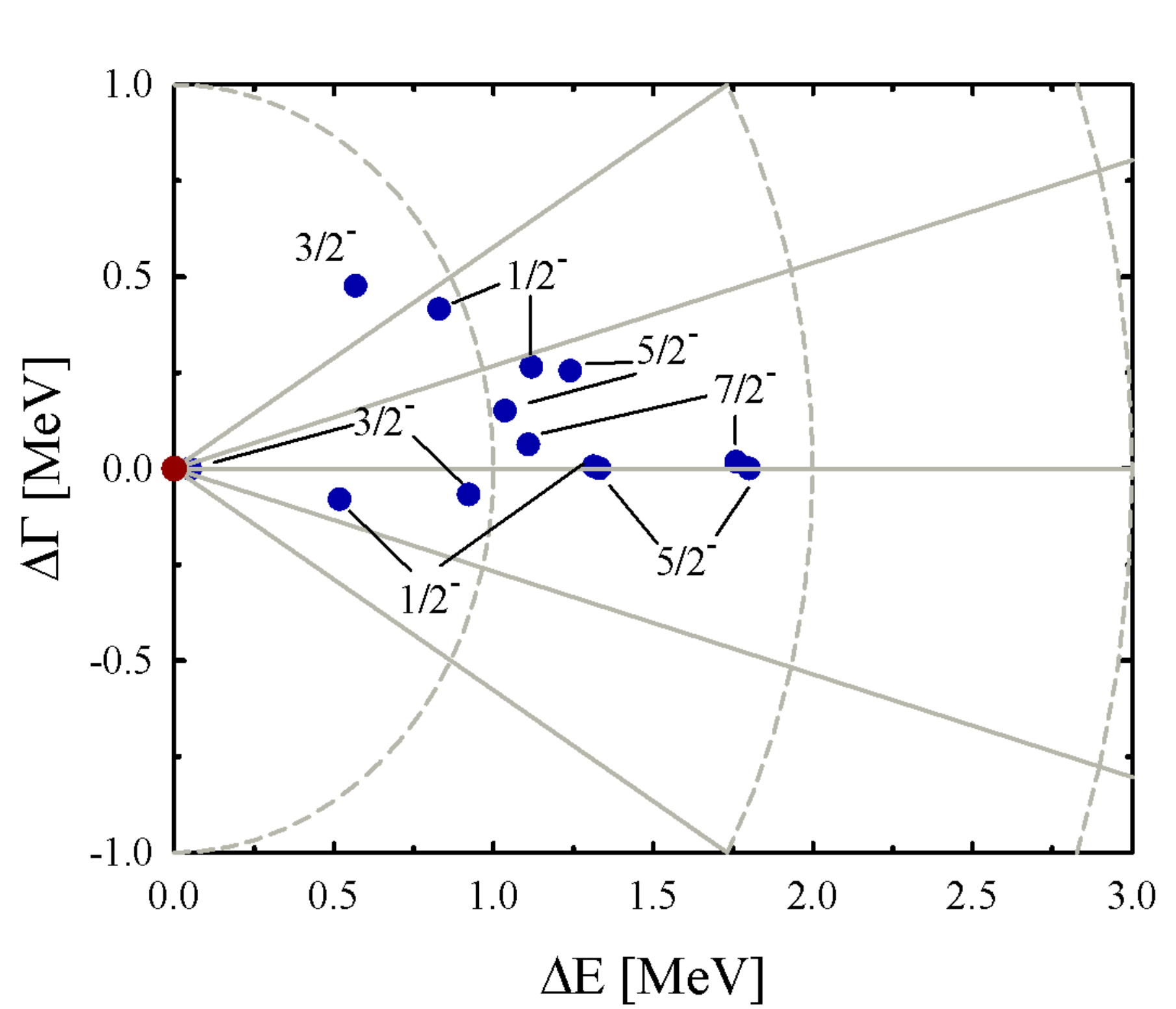}%
\caption{The shift and rotation of negative parity resonance states in $^{11}%
$C with respect to the position of their counterparts in $^{11}$B.}%
\label{Fig:DEvsDG11B11C}%
\end{center}
\end{figure}

\begin{figure}[ptbh]
\begin{center}
\includegraphics[width=\columnwidth]
{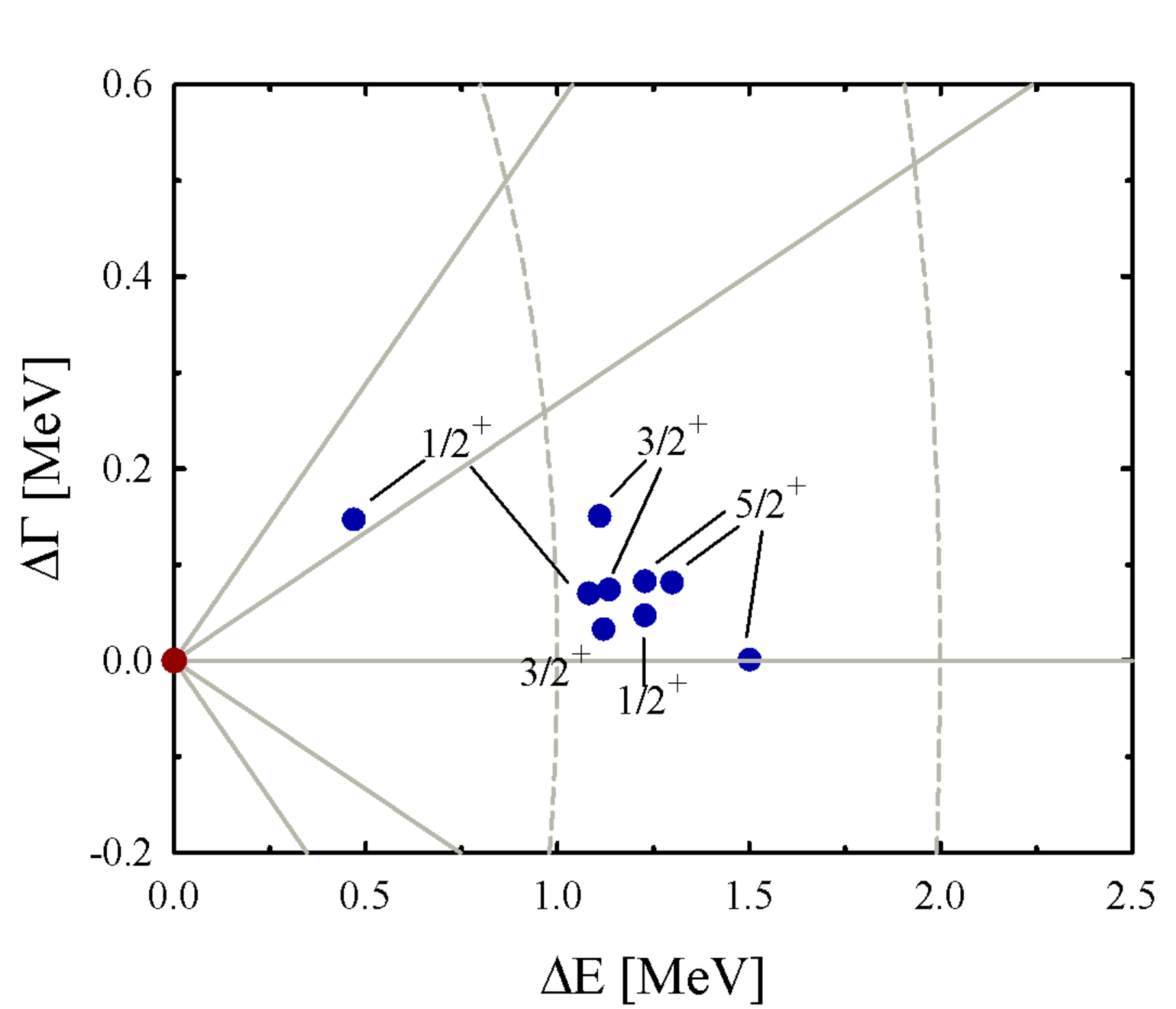}%
\caption{The Coulomb shift parameters for positive parity resonance states in 
$^{11}$B and $^{11}$C nuclei.}%
\label{Fig:DEvsDG11B11CP}%
\end{center}
\end{figure}

Effects of the Coulomb interaction on resonance states of the positive and
negative parity are demonstrated in Figs. \ref{Fig:Spect11B11CRSP} and
\ref{Fig:Spect11B11CRSN}, respectively. These figures demonstrate that for
the main part of resonance states in $^{11}$C have approximately the same
Coulomb shift  $R_{C}$ \ with respect to their counterparts in $^{11}$B. More
detail information about effects of the Coulomb interaction on resonance
states in the three-cluster continuum of $^{11}$B and $^{11}$C is presented in
Figs. \ref{Fig:DEvsDG11B11C}\ and \ref{Fig:DEvsDG11B11CP} .%

%

In Fig. \ref{Fig:DEvsDG11B11C}\ we show effects of the Coulomb interaction
on the negative parity resonance states in $^{11}$B and $^{11}$C nuclei. As
it can be seen, a large group of resonance states are concentrated around $%
R_{C}$=1 MeV. There are two resonance states with the Coulomb shift $%
R_{C}\approx $2 MeV. One of these resonance states, namely the 5/2$_{1}^{-}$
resonance state,  has a very small width and was regarded in Ref. \cite%
{2018PhRvC..98b4325V} as the Hoyle-analog states.  The majority of the
negative parity resonance states have small rotational angle 0$\leq \theta
_{C}\leq $18$^{\circ }$. Only two resonance states of 3/2$^{-}$ and 1/2$^{-}$ have the relatively
large Coulomb rotational angles $\theta _{C}\approx $37$^{\circ }$ and $%
\theta _{C}\approx $40$^{\circ }$. This is due to the fact that resonance
states in $^{11}$B and $^{11}$C nuclei have very small widths comparing to
the resonance states in $^{9}$Be and $^{9}$B. Thus one may conclude that the
Coulomb interaction has  moderate effects on the energies and widths of the
negative parity resonance states in $^{11}$B and $^{11}$C. 

What is new for resonance states in $^{11}$B and $^{11}$C? There are three
resonance states in $^{11}$C with the width smaller than their counterparts
in $^{11}$B. They are the 3/2$_{1}^{-}$, 3/2$_{2}^{-}$ and 1/2$_{6}^{-}$
resonance states. They have very small ($R_{C}\approx $0.05 MeV and $%
R_{C} \approx $0.5 MeV) \ or moderate ($R_{C} \approx $0.9 MeV) values of the
Coulomb shift. Thus for the resonance states of the negative parity in
mirror nuclei $^{11}$B and $^{11}$C, we observe the second scenario of
motion of resonance states caused by the Coulomb interaction.

For the positive parity states, the Coulomb parameters $R_{C}$ and $\theta_{C}$ are presented
in Fig. \ref{Fig:DEvsDG11B11CP}. As we can see, all these resonance states
belong to the first scenario, as both $\Delta E>0$ and $\Delta \Gamma >0$.
The main part of these resonance state are more  tightly concentrated around 
$R_{C}\approx $1 MeV then the negative parity states. Besides, the Coulomb
rotational angles $\theta _{C}$ for the positive parity states are also
smaller then for the negative parity states. They do not exceed 8$^{\circ }$.
There is one exception from this rule: the 1/2$_{1}^{+}$ resonance state
has a small value of the Coulomb shift $R_{C}\approx $0.5 MeV and relatively
large value of the Coulomb rotational angles $\theta _{C}\approx $17$^{\circ
}$. As for the negative parity states, the largest Coulomb shifts $%
R_{C}\approx $1.3 MeV and $R_{C}\approx $1.5 MeV are obtained for very
narrow resonance states 5/2$_{2}^{+}$ and 5/2$_{1}^{+}$, respectively. The
latter resonance state as was shown in Ref. \cite{2018PhRvC..98b4325V} is
the Hoyle-analog state.

Results presented in this section indicate that the Coulomb interaction
increases energy, and in many cases increases and in a few cases decreases width 
of resonance states in $^{11}$C. Thus
resonance states in the three-cluster continuum of the mirror nuclei $^{11}$B
and $^{11}$C realize the first and the second scenarios of motion of resonance states in the
\ $E-\Gamma$ plane.

\subsection{$^{7}$Li and $^{7}$Be}

Spectra of bound and resonance states in $^{7}$Li and $^{7}$Be have been
calculated within AMGOB. In this model, three-cluster configurations,
specified in Table \ref{Tab:MirrorNuclei}, were projected onto a set of the
two-body channels $^{4}$He+$^{3}$H and $^{6}$Li+$n$ in $^{7}$Li and $^{4}%
$He+$^{3}$He and $^{6}$Li+$p$ in $^{7}$Be. These are the dominant two-body
channels of $^{7}$Li and $^{7}$Be. The the AMGOB model also allowed us to
study polarizability of interacting clusters when they approach each other. It
was shown in Refs. \cite{2009PAN....72.1450N, 2009NuPhA.824...37V,
2017UkrJPh..62..461V} that the polarization of interacting clusters
substantially decreases energy and width of resonance states in a compound nucleus.

In Table \ref{Tab:Spect7Li7Be} we collect the energy of bound states and the
energy and width of resonance states in the mirror nuclei $^{7}$Li and $^{7}%
$Be. These quantities were calculated with the MP in Refs.
\cite{2009PAN....72.1450N, 2009NuPhA.824...37V}. Note that the
7/2$^{-}$ resonance states presented in Table \ref{Tab:Spect7Li7Be} reside in
the $^{4}$He+$^{3}$H and $^{4}$He+$^{3}$He two-body continuum, while the
5/2$^{-}$ resonance states belong to the energy region where there are two
open channels $^{4}$He+$^{3}$H and $^{6}$Li+$n$ in $^{7}$Li, and $^{4}%
$He+$^{3}$He and $^{6}$Li+$p$ in $^{7}$Be.%

\begin{table}[tbp] \centering
\caption{Spectra of bound and resonance states in $^7$Li and $^7$Be. Energies and widths are in MeV.}%
\begin{tabular}
[c]{ccccccc}\hline
& \multicolumn{2}{c}{$^{7}$Li} & \multicolumn{2}{c}{$^{7}$Be} &
\multicolumn{2}{c}{Coulomb shifts}\\\hline
$J^{\pi}$ & $E$ & $\Gamma$ & $E$ & $\Gamma$ & $R_{C}$ & $\theta_{C}$\\\hline
$3/2_{1}^{-}$ & -2.721 & 0.0 & -1.702 & 0.0 & 1.019 & 0\\
$1/2_{1}^{-}$ & -2.469 & 0.0 & -1.412 & 0.0 & 1.057 & 0\\
$7/2_{1}^{-}$ & 2.052 & 0.073 & 2.820 & 0.130 & 0.770 & 4.24\\
$5/2_{1}^{-}$ & 4.270 & 1.104 & 5.040 & 1.343 & 0.806 & 17.24\\\hline
\end{tabular}
\label{Tab:Spect7Li7Be}%
\end{table}%

\begin{figure}[ptbh]
\begin{center}
\includegraphics[width=\columnwidth]
{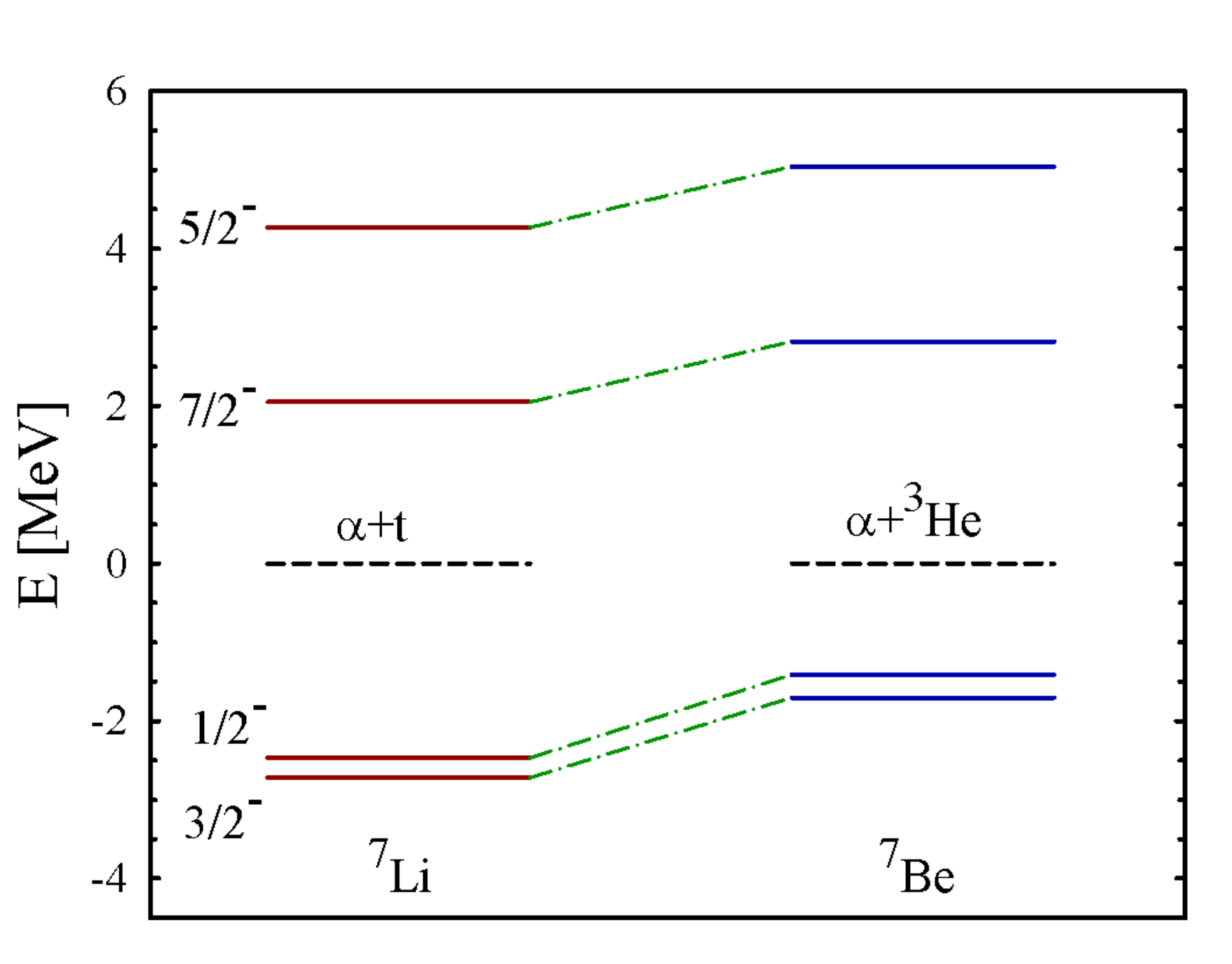}%
\caption{Spectrum of bound and resonance states in $^{7}$Li and $^{7}$Be.}%
\label{Fig:Spectr7Li7Be}%
\end{center}
\end{figure}

\begin{figure}[ptbh]
\begin{center}
\includegraphics[width=\columnwidth]
{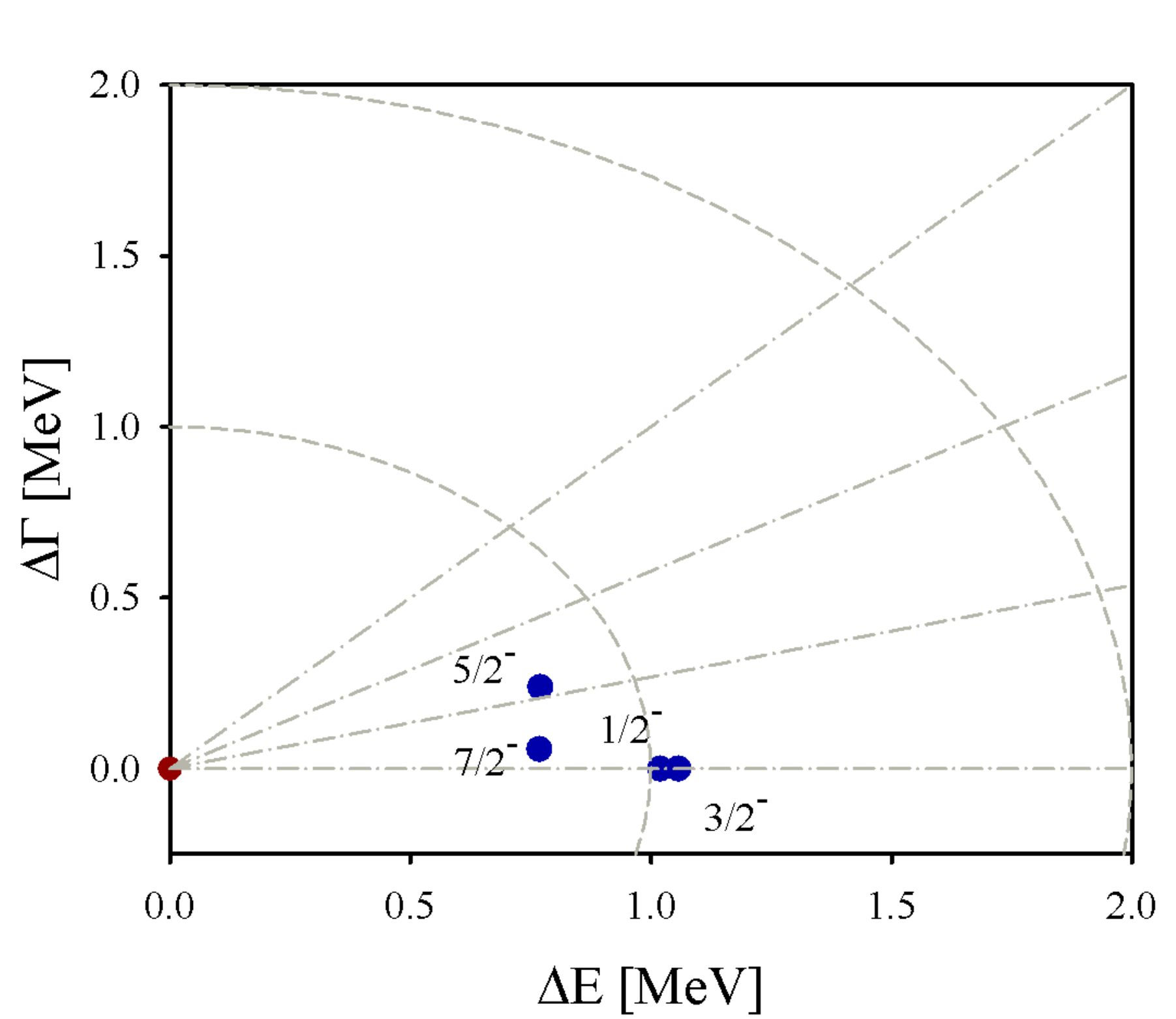}%
\caption{The Coulomb shift of bound and resonance states in $^{7}$Be with
respect to corresponding states in $^{7}$Li.}%
\label{Fig:DEvsDG7Li7Be}%
\end{center}
\end{figure}

In Fig. \ref{Fig:Spectr7Li7Be} we show the relative position of bound and
resonance states in $^{7}$Li and $^{7}$Be, and in Fig. \ref{Fig:DEvsDG7Li7Be}
we\ display their Coulomb shift and rotation.%

These figures and Table \ref{Tab:Spect7Li7Be} show that the stronger Coulomb
interaction in $^{7}$Be shifts all bound and resonance states (with respect to
their position in $^{7}$Li) approximately on the same value and rotate the
broad 5/2$^{-}$ resonance state on 17 degrees while it rotates the narrow
7/2$^{-}$ state on 4 degrees.

\subsection{$^{8}$Li and $^{8}$B}

Let us now consider how the Coulomb interaction affects the spectra of bound
and resonance states in mirror nuclei $^{8}$Li and $^{8}$B. These nuclei
similar to nuclei $^{7}$Li and $^{7}$Be were studied within the AMGOB in Ref.
\cite{2017UkrJPh..62..461V}. The three-cluster configurations $\alpha+t+n$ and
$\alpha+^{3}$He$+p$ were projected on the dominant the two-cluster channels
$^{7}$Li$+n$ and $^{7}$Be$+p$. \ We restricted ourselves with a single-channel
approximation in an asymptotic region of $^{8}$Li and $^{8}$B, as bound states
exist only in two-cluster subsystems $\alpha+t$ and $\alpha+^{3}$He. Thus
resonance states in $^{8}$Li and $^{8}$B, which are displayed in Table
\ref{Tab:Spect8Livs8B} together with bound states, belong to the two-body
continua $^{7}$Li$+n$ and $^{7}$Be$+p$, respectively.

As we see in Table \ref{Tab:Spect8Livs8B}, the Coulomb interaction diminished
number of bound states in $^{8}$B with respect to $^{8}$Li. Thus, the
effective interaction between clusters is reduced by the Coulomb interaction,
\ and it results in decreasing (diminishing) energy of the 2$^{+}$ ground
state \ and moving up the 1$^{+}$ excited state to continuous spectrum  (i.e.
transforming the 1$^{+}$ bound state into a resonance state).%

\begin{table}[tbp] \centering
\caption{Spectra of bound and resonance states in $^8$Li and $^8$B 
and the Coulomb shift parameters. Results are obtained with the MHNP. 
Energies and widths are in MeV.}%
\begin{tabular}
[c]{ccccccc}\hline
& \multicolumn{2}{c}{$^{8}$Li} & \multicolumn{2}{c}{$^{8}$B} &
\multicolumn{2}{c}{Coulomb shift}\\\hline
$J^{\pi}$ & $E$ & $\Gamma$ & $E$ & $\Gamma$ & $R_{C}$ & $\theta_{C}$\\\hline
$2_{1}^{+}$ & -1.908 & 0.0 & -0.1393 & 0.0 & 1.769 & 0.00\\
$1_{1}^{+}$ & -0.977 & 0.0 & 0.615 & 0.044 & 1.592 & 1.57\\
$3_{1}^{+}$ & 0.610 & 0.165 & 2.560 & 0.572 & 1.992 & 11.79\\
$1_{2}^{+}$ & 0.014 & 0.002 & 1.305 & 0.600 & 1.423 & 24.86\\
$1_{3}^{+}$ & 1.002 & 1.433 & 3.218 & 2.089 & 2.311 & 16.49\\
$1_{4}^{+}$ & 2.129 & 0.913 & 3.916 & 0.272 & 1.898 & -19.71\\
$2_{2}^{+}$ & 1.436 & 0.658 & 3.321 & 1.139 & 1.945 & 14.31\\
$2_{3}^{+}$ & 3.175 & 0.976 & 3.889 & 0.346 & 0.953 & -41.45\\
$4_{1}^{+}$ & 3.190 & 0.002 & 4.226 & 0.775 & 1.293 & 36.74\\
$2_{1}^{-}$ & 3.494 & 0.365 & 3.747 & 0.712 & 0.430 & 53.92\\
$1_{1}^{-}$ & 0.681 & 0.6245 & 1.132 & 1.828 & 1.285 & 69.45\\
$3_{1}^{-}$ & 3.756 & 0.883 & 3.957 & 1.495 & 0.644 & 71.83\\\hline
\end{tabular}
\label{Tab:Spect8Livs8B}%
\end{table}%

\begin{figure}[ptbh]
\begin{center}
\includegraphics[width=\columnwidth]
{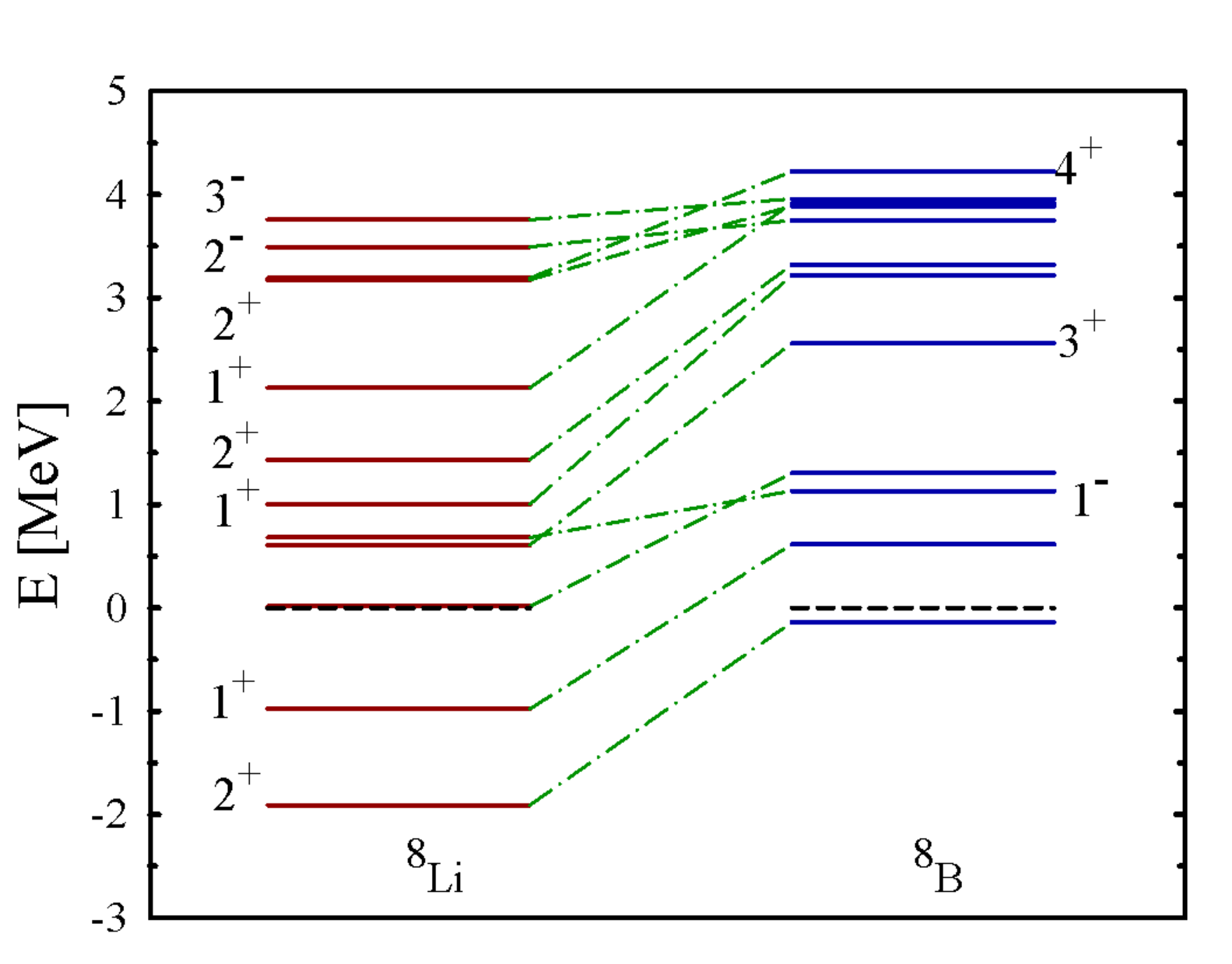}%
\caption{Effects of the Coulomb forces on position of resonance states in 
$^{8}Li$ and $^{8}B$.}%
\label{Fig:Spectr8Li8B}%
\end{center}
\end{figure}

More interesting and intriguing is the influence of the Coulomb forces on
energy and width of resonance states. As was shown in Ref.
\cite{2005PAN....68.1147T} effects of the Coulomb forces on resonance states
even in two-cluster systems are not trivial. Here we deal with three-cluster
system projected onto a set of two-cluster channels. In Fig.
\ref{Fig:Spectr8Li8B} we compare spectrum of bound and resonance states in
$^{8}$Li and $^{8}$B calculated with the MHNP. Dot-dashed lines connect states
with the same value of the total angular momentum $J$ and parity $\pi$. We can
see that the Coulomb interaction shifted up energy of all bound and resonance
states. Effects of the Coulomb interaction are the same for all states  except
for $3^{+}$ and $2^{-}$ resonance states. As we can see the $2^{-}$ state has
the smallest impact of the Coulomb interaction on energy of this state, while
the largest impact is observed for the $3^{+}$ resonance state. The main
result of our consideration is that the Coulomb forces substantially increase
width of resonance states in $^{8}$B with respect to the corresponding
resonance states in $^{8}$Li.%

\begin{figure}[ptb]
\begin{center}
\includegraphics[width=\columnwidth]
{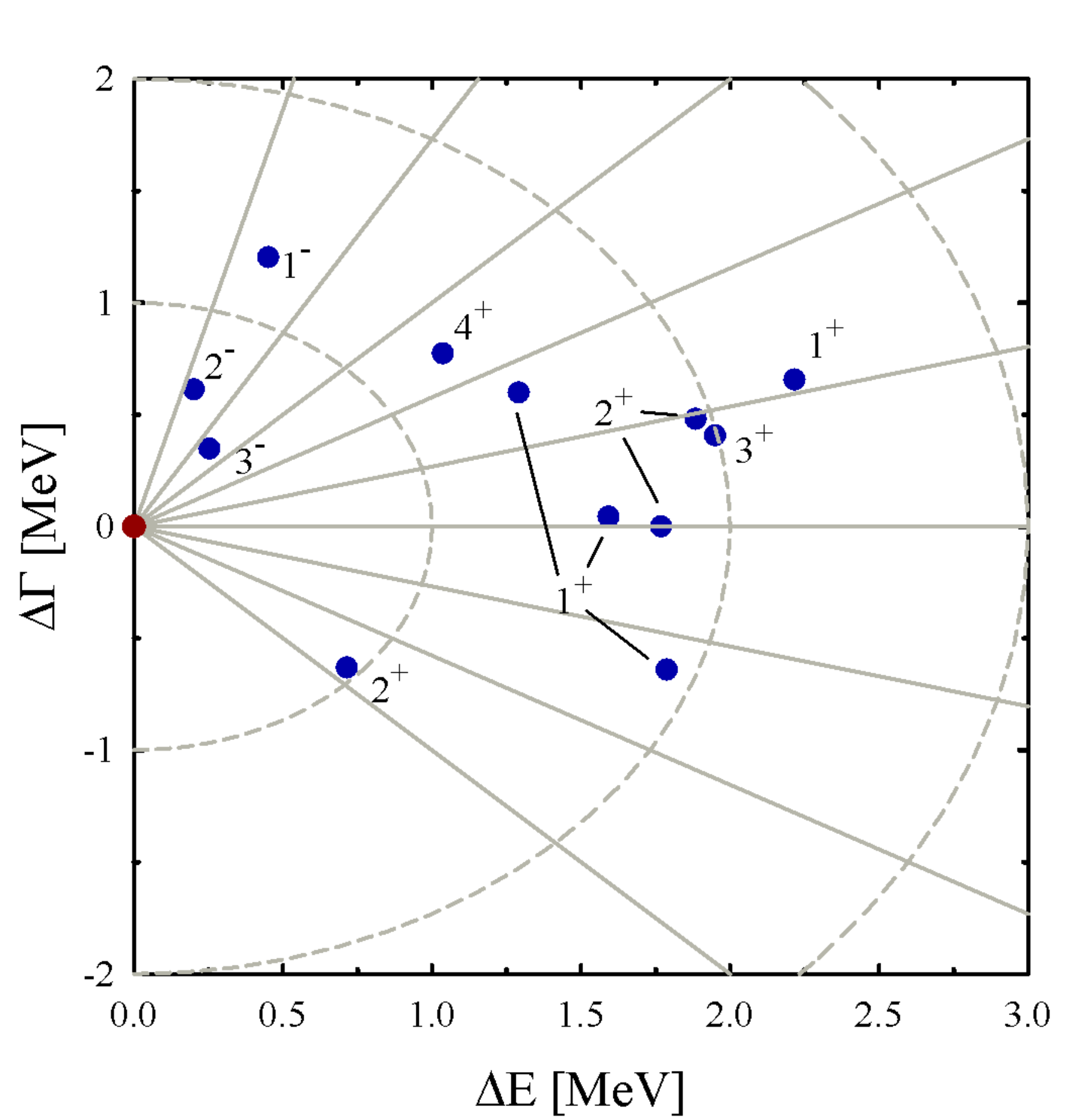}%
\caption{The shift and rotation of parameters of resonance states in mirror
nuclei $^{8}$Li and $^{8}$B caused by the Coulomb interaction.}%
\label{Fig:DEvsDG8Li8B}%
\end{center}
\end{figure}

More detail information about effects of the Coulomb interaction in mirror
nuclei $^{8}$Li and $^{8}$B are displayed in Fig. \ref{Fig:DEvsDG8Li8B}. In
this case we can also distinguish resonance states with the strong effects,
and they are located around $R_{C}\approx$ 2 MeV, with the medium effects they
are close to $R_{C}\approx$ 2 MeV, and resonance states with the small effects
which have $R_{C}\approx$ 0.5 MeV. The relative position of resonance states
in $^{8}$B with respect to their counterparts in $^{8}$Li shows that there are
three resonance states ($1_{3}^{+}$, $1_{4}^{+}$ and $2_{3}^{+}$) with
negative values of $\Delta\Gamma$, thus in this pair of nuclei we observe the
second scenario. The Coulomb interaction decreases the width of two resonance
states in $^{8}$B, but increases their energy. It is worthwhile to recall that
resonance states $^{8}$Li and $^{8}$B are determined in the two-body continuum.%

\section{Analysis of resonance wave functions}
\label{Sec:WaveFuns}

In this section we consider how the Coulomb forces affect wave functions of
resonance states in the mirror nuclei. These effects are considered for the
nuclei $^{9}$Be and $^{9}$B, $^{11}$B and $^{11}$C as they are common for all
nuclei under consideration. Besides, two scenarios are realized in the $^{11}%
$B and $^{11}$C nuclei. It is then interesting to study wave functions of
resonance states when (i) the Coulomb forces increase the energy and width of
resonance state, and (ii) when they increase the energy but decrease the width
of resonance state. A wave function of many-channel system is a very
complicated many-dimension object which is difficult to analyze. Thus we
present wave functions of bound and resonance states of three-cluster systems
in a compressed form through the weights $W_{sh}$ of oscillator shell $N_{sh}$
($N_{sh}$=0, 1, 2,\ldots) in the corresponding wave function. 
The weights suggest a
spectral decomposition of a three-cluster wave function in terms of the
probability (for bound states) or contribution (for scattering states) of
many-body oscillator functions belonging to the oscillator shell $N_{sh}$. It
is important to recall that the oscillator functions with small values of
$N_{sh}$ \ describe a compact configuration of three-cluster system, while the
oscillator functions with large values of $N_{sh}$ \ reproduce a dispersed
three-cluster configuration. It is also important to recall that wave
functions of discrete and continuous spectrum states are properly normalized
(see Eqs. (23) and (24) in Ref. \cite{2018PhRvC..98b4325V}). As in Refs.
\cite{2017PhRvC..96c4322V, 2018PhRvC..98b4325V} we will display only
the internal part of a wave function of resonance states.

\begin{figure}[ptb]
\begin{center}
\includegraphics[width=\columnwidth]
{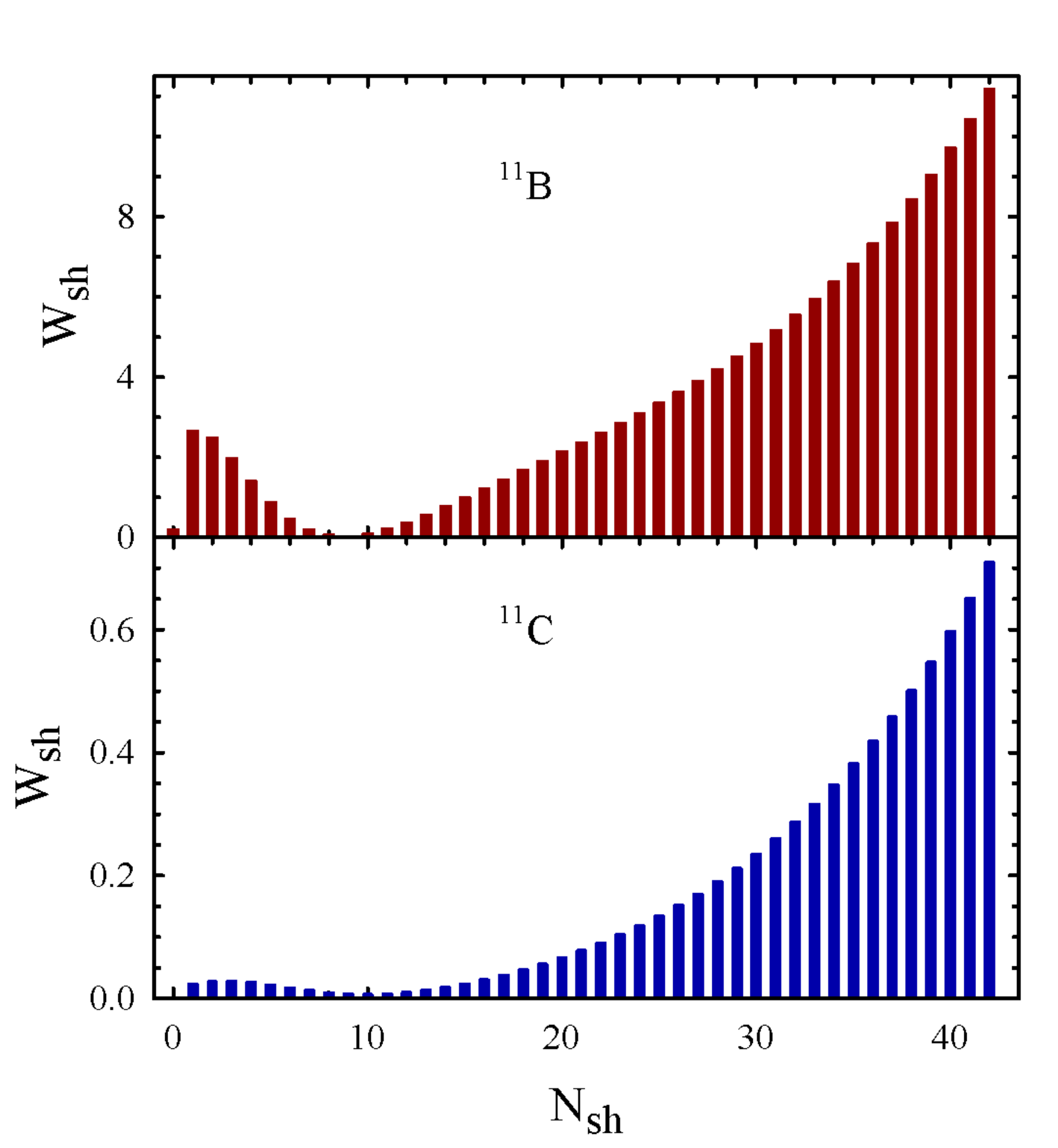}%
\caption{Weights of different oscillator shells $W_{sh}$ as a function of
$N_{sh}$ in the wave functions of the 1/2$^{+}_{1}$ resonance state in $^{11}$B
and $^{11}$C.}%
\label{Fig:ShWeight12pRS11B11C}%
\end{center}
\end{figure}

We start with the case of resonance states for which we observed the smallest
impact of the Coulomb forces. These are the 1/2$^{+}$ resonance states in
$^{11}$B and $^{11}$C. In Fig. \ref{Fig:ShWeight12pRS11B11C} we compare the
shell weights for the first 1/2$^{+}$ resonance state in $^{11}$B and $^{11}%
$C. We can see they have approximately the same shape, however the amplitude
of $W_{sh}$ for $^{11}$B is 15 times larger than these weights for $^{11}$C.
This is the effect of the Coulomb forces. Fig. \ref{Fig:ShWeight12pRS11B11C}
helps us to explain why the effects of the Coulomb forces on the energy and
width of the 1/2$^{+}$ resonance state is small. We see that the weights of
compact three-cluster configuration are negligible small for both nuclei, and
only dispersed configurations dominate. For them the Coulomb interaction is
not so strong as for compact configurations. These conclusions are valid for
the 1/2$^{+}_{1}$ resonance states in the mirror nuclei $^{9}$Be and $^{9}$B (see
Fig. 5 in Ref. \cite{2018PhRvC..98b4325V}). In this case, the effects of the
Coulomb interaction are not so strong as in $^{11}$B and $^{11}$C, however the
shapes of the resonance wave functions are very similar.

\begin{figure}[ptb]
\begin{center}
\includegraphics[width=\columnwidth]
{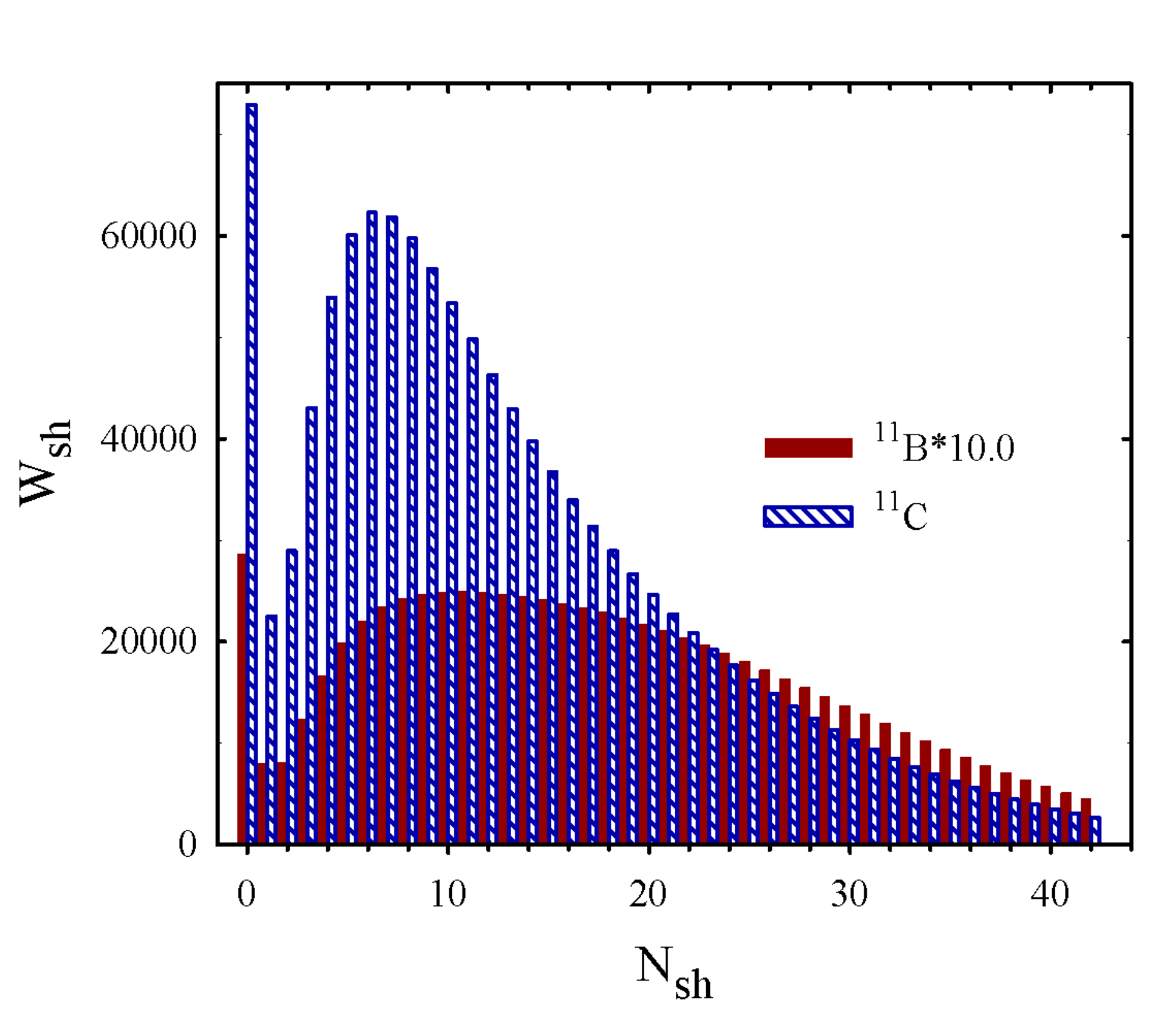}%
\caption{Weights $W_{sh}$ of different oscillator shells  in the wave
functions of the 3/2$^{-}_{1}$ resonance states in $^{11}$B and $^{11}$C.}%
\label{Fig:SW32MRS11B11CR1}%
\end{center}

\end{figure}

\begin{figure}[ptb]
\begin{center}
\includegraphics[width=\columnwidth]
{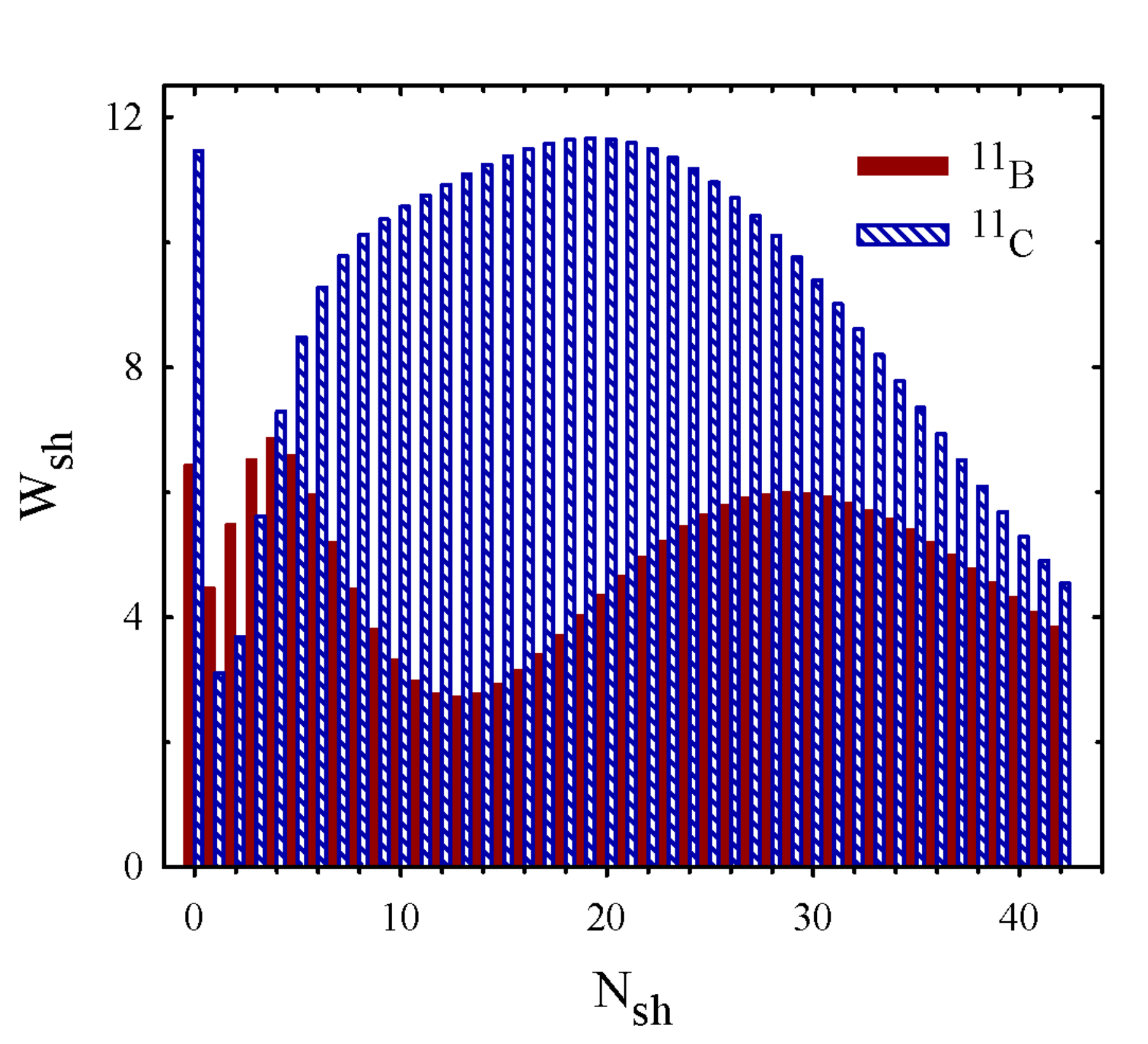}%
\caption{Weights $W_{sh}$ of different oscillator shells  in the wave
functions of the 3/2$^{-}_{2}$ resonance states in $^{11}$B and $^{11}$C.}%
\label{Fig:SW32MRS11B11CR2}%
\end{center}
\end{figure}

\begin{figure}[ptb]
\begin{center}
\includegraphics[width=\columnwidth]
{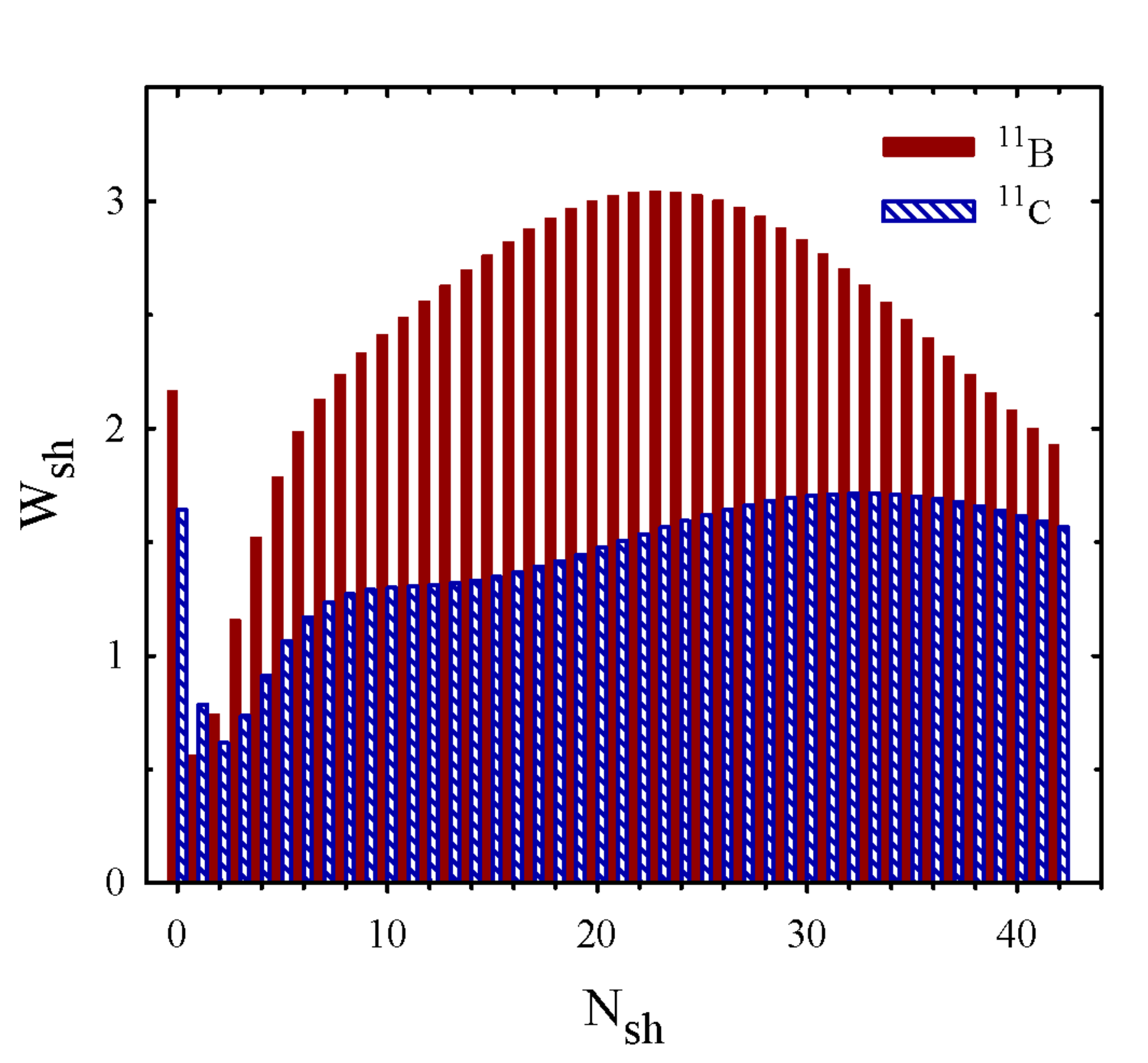}%
\caption{Weights $W_{sh}$ of different oscillator shells  in the wave
functions of the 3/2$^{-}_{3}$ resonance states in $^{11}$B and $^{11}$C.}%
\label{Fig:SW32MRS11B11CR3}%
\end{center}
\end{figure}

Let us now turn our attention to the 3/2$_{1}^{-}$, 3/2$_{2}^{-}$ and
3/2$_{3}^{-}$ resonance states in $^{11}$B and $^{11}$C. \ As we can see from
Table \ref{Tab:Spect11Bvs11C}, the Coulomb interaction realizes two scenarios
for these resonance states. The second scenario is observed for the
3/2$_{1}^{-}$and 3/2$_{2}^{-}$ resonance states, and the first scenario is
realized for the 3/2$_{3}^{-}$ resonance state. Thus we consider wave
functions of these three-resonance states. In Figs. \ref{Fig:SW32MRS11B11CR1}, 
\ref{Fig:SW32MRS11B11CR2} and \ref{Fig:SW32MRS11B11CR3}
we display the weights of the oscillator
shells for the 3/2$^{-}_{1}, $3/2$^{-}_{2}$ and 3/2$^{-}_{3}$ resonance states, respectively. 
The resonance states 3/2$_{1}^{-}$, 3/2$_{2}^{-}$ and 3/2$_{3}^{-}$ are denotes as
$R_{1}$, $R_{2}$ and $R_{3}$, respectively. The $R_{1}$ states are very narrow
resonance states in $^{11}$B and $^{11}$C. The Coulomb interaction
makes this resonance state in $^{11}$C more narrow than in $^{11}$B, as the
result the amplitude of the $^{11}$C wave function is approximately 25 larger
than the amplitude of the $^{11}$B wave function. Thus the stronger Coulomb
interaction in $^{11}$C does not decrease the amplitude of the wave function, as
was observed in other cases demonstrated above or in Refs.
\cite{2013UkrJPh.58.544V, 2017PhRvC..96c4322V,
2018PhRvC..98b4325V}. On the contrary, it substantially increases the
amplitude of the wave function of such a narrow resonance state. There is an
interplay of two factors which affects the wave function of the resonance state.
First, the stronger Coulomb interaction suppresses stronger a wave functions
of  resonance states at small distance between interacting clusters, or, in our
representation, oscillator shells with small values of $N_{sh}$.  Second,
the narrower is the resonance state, the larger is the amplitude of the
resonance wave function. Thus, in the $R_{1}$ resonance state the second
factor dominates over the first factor and as a results  increases the
amplitude of the wave function of $^{11}$C in the internal region. 

The $R_{2}%
$\ resonance states are not so narrow as the $R_{1}$\ resonance states. These
resonance states realize the second scenario when the Coulomb interaction
decreases the width but increases the energy of the resonance state. As in
previous case, the Coulomb interaction increases the amplitude of the $R_{2}$
wave function in $^{11}$C comparatively to the wave function of $^{11}$B,
however only two times. The $R_{3}$ resonance states realizes the first
scenario. In this case, the Coulomb interaction yields the moderate shift
$R_{C}$= 0.741 MeV but the large rotational angle $\theta_{C}$=39.96$^{\circ}%
$. The amplitude of the $R_{3}$ wave function in $^{11}$B is slightly larger
than the amplitude of the $R_{3}$ wave function in $^{11}$C.%


To gain more information about structure of resonance wave functions which are
related by the first and second scenarios, we consider the 1/2$_{3}^{-}$,
1/2$_{4}^{-}$ and 1/2$_{5}^{-}$ resonance states in $^{11}$B and $^{11}$C. We
do not consider the 1/2$_{3}^{-}$ states as being a weakly bound state in
$^{11}$B it transformed into a resonance state in $^{11}$C.

\begin{figure}[ptb]
\begin{center}
\includegraphics[width=\columnwidth]
{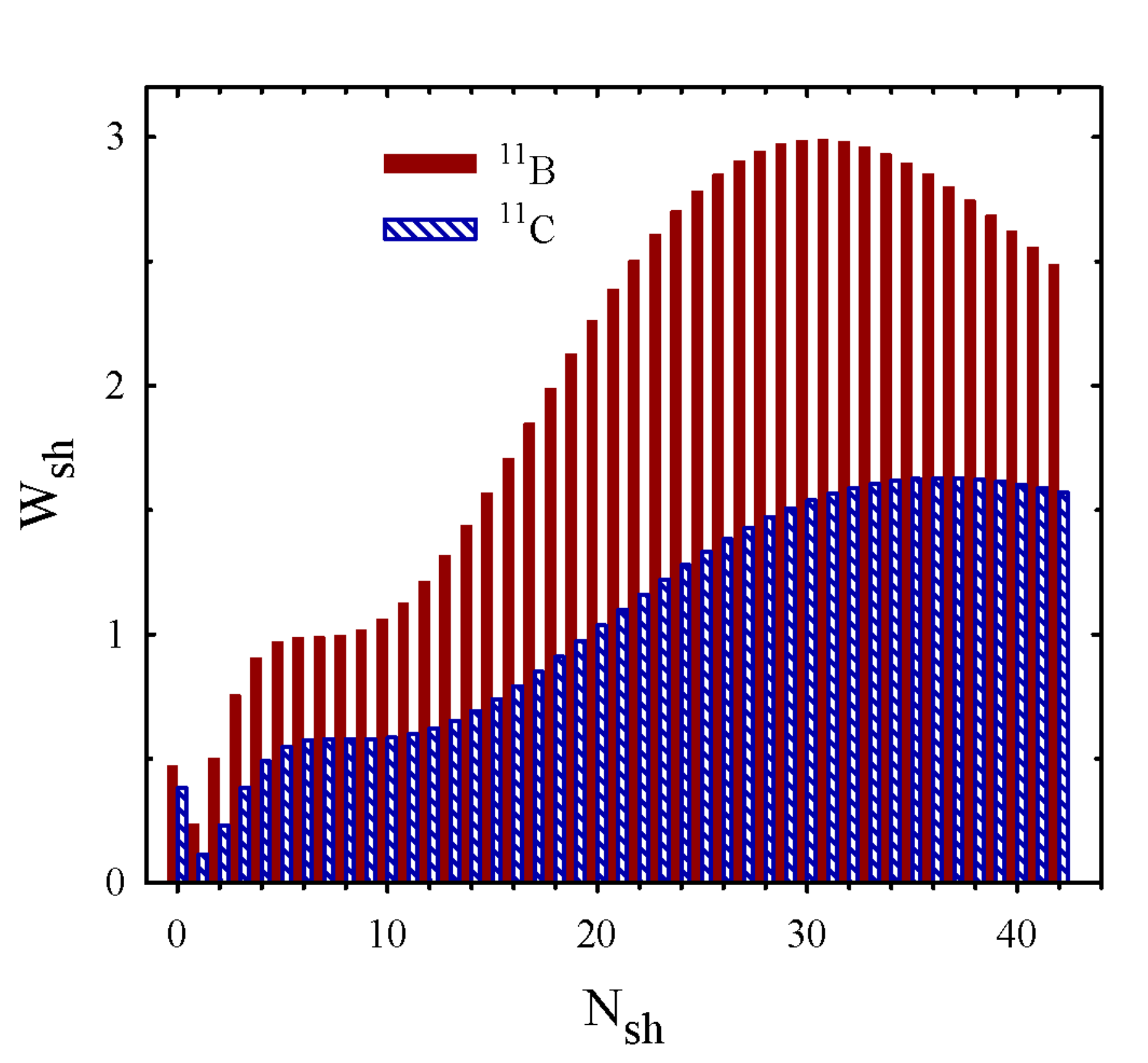}%
\caption{The weights of different oscillator shells in the wave functions of
the 1/2$_{4}^{-}$ resonance states in mirror nuclei $^{11}$B and $^{11}$C}%
\label{Fig:SW12MRS11B11CR1}%
\end{center}
\end{figure}

\begin{figure}[ptb]
\begin{center}
\includegraphics[width=\columnwidth]
{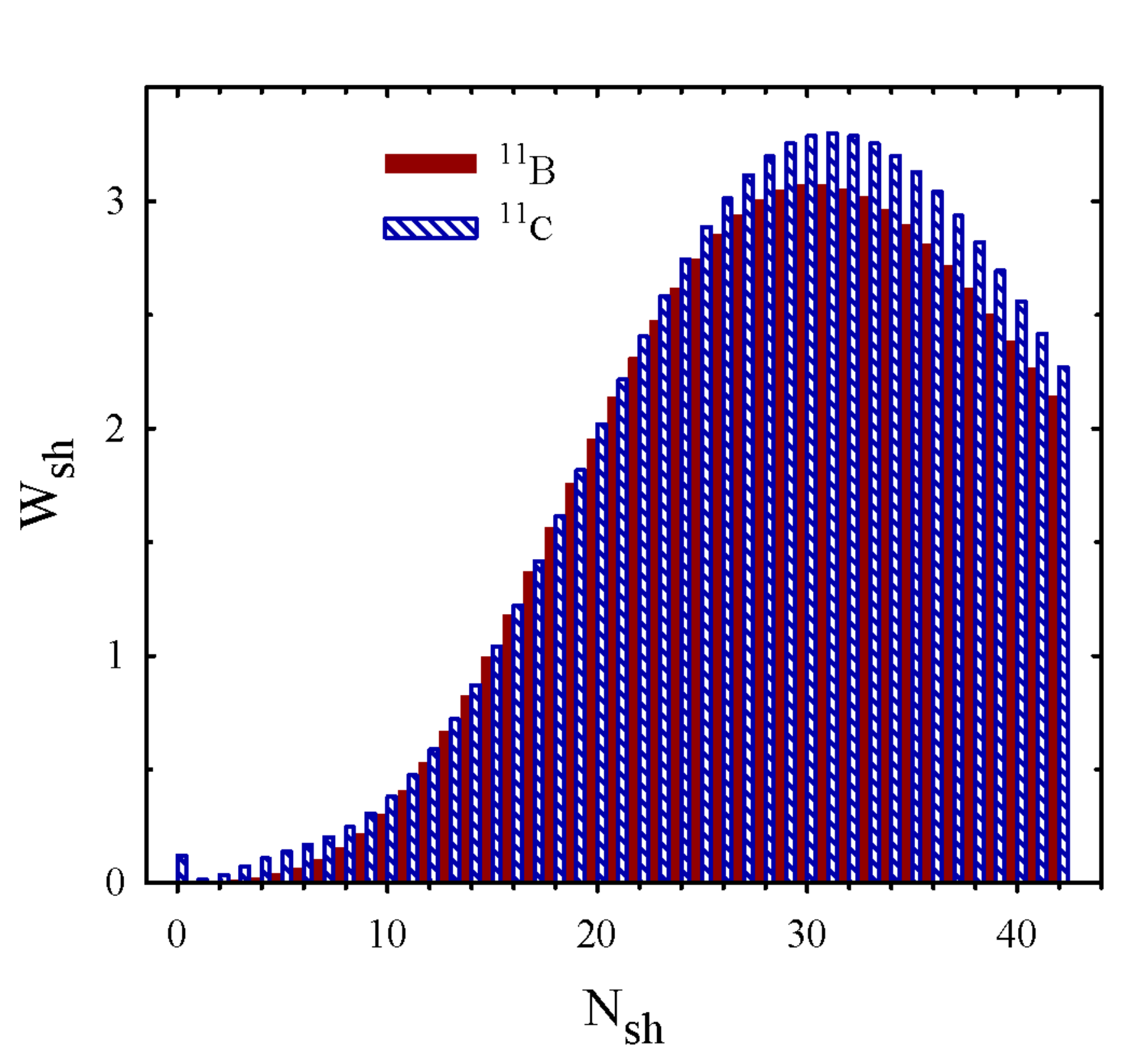}%
\caption{The structure of the wave functions of the 1/2$_{6}^{-}$ resonance
states in $^{11}$B and $^{11}$C}%
\label{Fig:SW12MRS11B11CR3}%
\end{center}
\end{figure}

Structure of wave functions of  two 1/2$^{-}$ resonance states in $^{11}$B and
$^{11}$C are presented in Figs. \ref{Fig:SW12MRS11B11CR1} and
\ref{Fig:SW12MRS11B11CR3} in terms of the weights $W_{sh}$ of different
oscillator shells. We selected \ wave functions of the 1/2$_{4}^{-}$ and
1/2$_{6}^{-}$ resonance states. The 1/2$_{4}^{-}$ resonance states are related
by the first scenario and we see that the Coulomb interaction suppress the
wave function in $^{11}$C with respect to the one in $^{11}$B. The same
results are also obtained for the 1/2$_{5}^{-}$ resonance states which are
united the first scenario. The 1/2$_{6}^{-}$ resonance states represents the
second scenario and in this case the wave function of the resonance state in
$^{11}$C at small distance (or at small values of $N_{sh}$) is slightly larger
than in $^{11}$B. Comparing Fig. \ref{Fig:SW12MRS11B11CR3} with Figs.
\ref{Fig:SW32MRS11B11CR1}, \ref{Fig:SW32MRS11B11CR2}, we came to the
conclusion that for resonance states related by the second scenario the
smaller is the width of resonance state in the $^{11}$C nucleus the larger is
the resonance wave function in the internal region for the $^{11}$C nucleus
with respect to one in the $^{11}$B nucleus. 

\begin{figure}[ptb]
\begin{center}
\includegraphics[width=\columnwidth]
{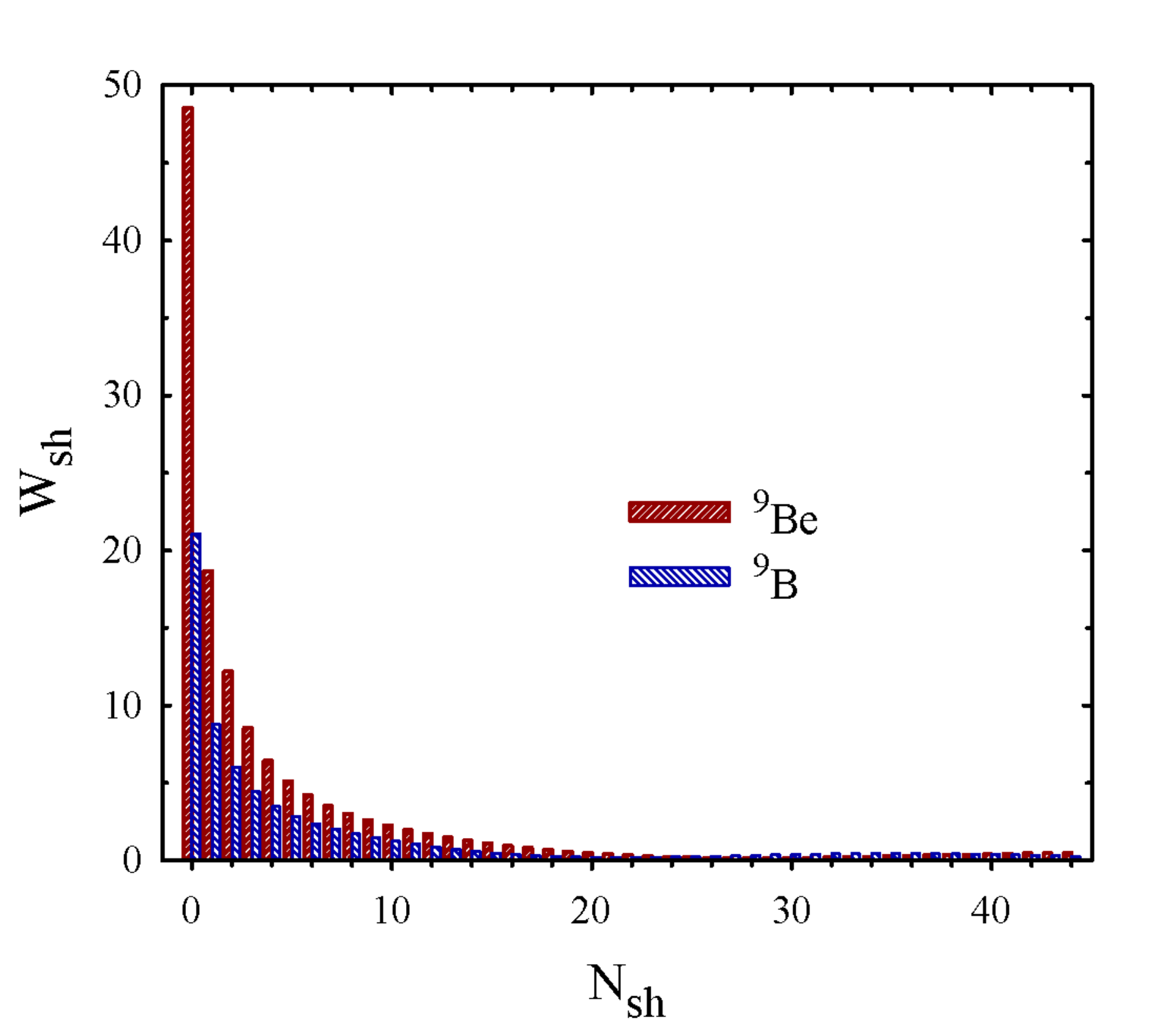}%
\caption{The weights of the oscillator shells in the wave functions of the
7/2$^{-}$ resonance states in $^{9}$Be and $^{9}$B.}%
\label{Fig:SW72MRS9Be9B}%
\end{center}
\end{figure}

It is interesting to consider the case when the Coulomb interaction has the
largest impact on the parameters of resonance states. For this aim we selected
the 7/2$^{-}$ and 3/2$^{+}$ resonance states in mirror nuclei $^{9}$Be and
$^{9}$B. First, we consider the 7/2$^{-}$ resonance states. The Coulomb
interaction shifts the 7/2$^{-}$ resonance state on $R_{C}$=2.072 MeV and
rotates on $\theta_{C}$=13.74$^{\circ}$. Fig. \ref{Fig:SW72MRS9Be9B} displays
wave functions of the 7/2$^{-}$ resonance states. As we see, wave functions of
these resonance states describe compact three-cluster configurations since the
oscillator shells with small values of $N_{sh}$ (0$\leq N_{sh}<$%
10) give the main contribution to these wave functions. The Coulomb
interaction reduces the amplitude of $W_{sh}$ of the $^{9}$B wave function
approximately two times with respect to the wave function of $^{9}$Be.%

\begin{figure}[ptb]
\begin{center}
\includegraphics[width=\columnwidth]
{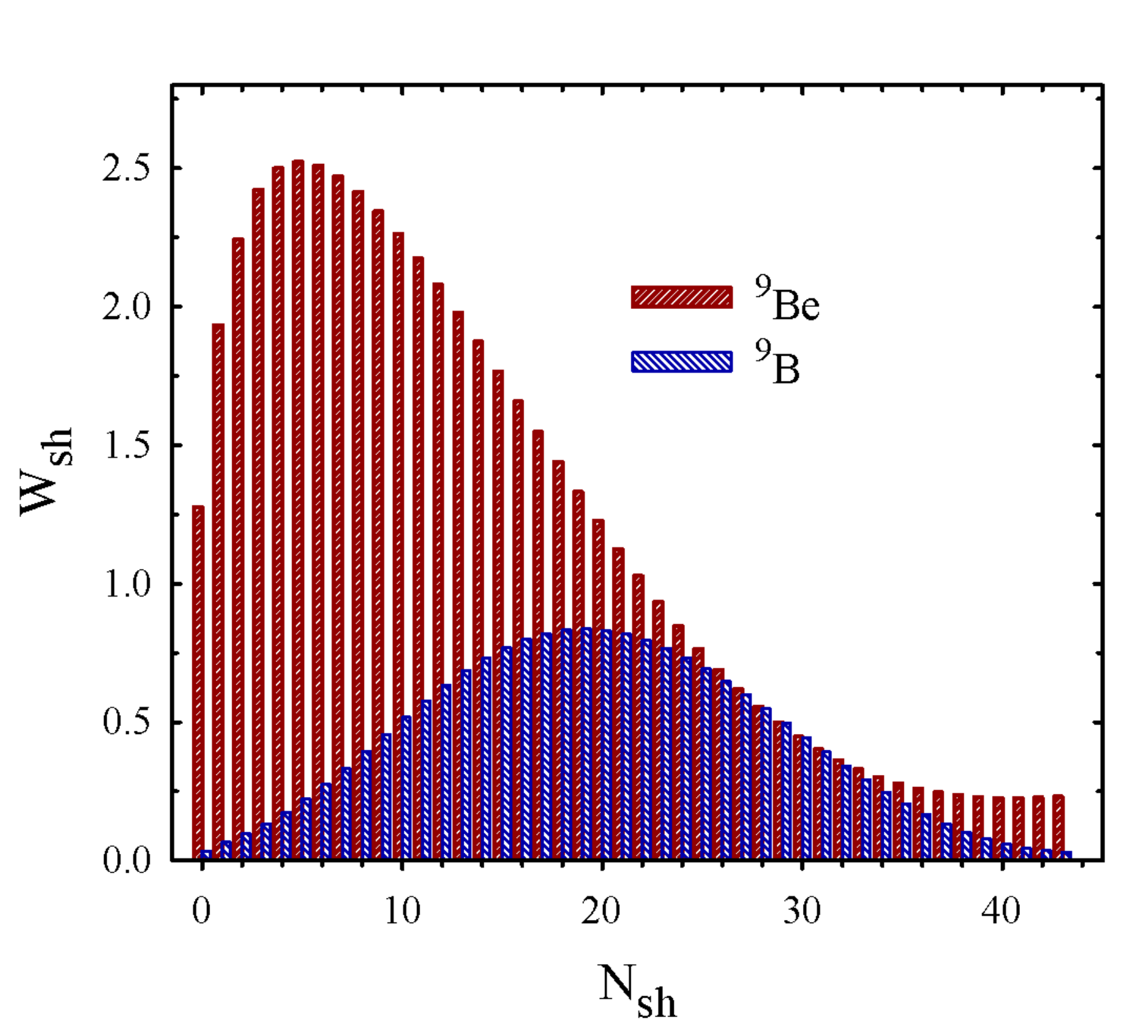}%
\caption{The weights of oscillator shells in wave functions of the 3/2$^{+}$
resonance states in $^{9}$Be and $^{9}$B.}%
\label{Fig:SW32PRS9Be9B}%
\end{center}
\end{figure}

The 3/2$^{+}$ resonance states in $^{9}$Be and $^{9}$B are related by the
largest Coulomb shift $R_{C}$=2.669 MeV and the largest Coulomb rotational
angle $\theta_{C}$=83.44$^{\circ}$. The weights of the oscillator shells
displayed in Fig. \ref{Fig:SW32PRS9Be9B} demonstrate that in this case the
Coulomb interaction dramatically changes the structure of the resonance wave
function in $^{9}$B. It strongly diminishes the weights of the oscillator
shells with small values of $N_{sh}$ (0$\leq N_{sh}\leq$15) and substantially
suppress the weights in the region 15$<N_{sh}\leq$24.%

In Figs. \ref{Fig:ShWeight12pRS11B11C}, \ref{Fig:SW32MRS11B11CR1},
\ref{Fig:SW32MRS11B11CR2}, \ref{Fig:SW32MRS11B11CR3}, \ref{Fig:SW72MRS9Be9B}, 
\ref{Fig:SW32PRS9Be9B} we
display behavior of the resonance wave functions which are typical for (or similar
to) all other resonance states and nuclei considered.

\section{Conclusions \label{Sec:Conclusions}}

We have considered effects of the Coulomb interaction on energies and widths
of resonance states in mirror nuclei $^{7}$Li and $^{7}$Be, $^{8}$Li and
$^{8}$B, $^{9}$Be and $^{9}$B, $^{11}$B and $^{11}$C. We have analyzed
resonance states embedded in two- and three-cluster continua of these nuclei.
Resonance states were obtained in the framework of  microscopic three-cluster
models. For the proper investigation of effects of the Coulomb interaction we
introduced two parameters which determine a rotation and shift (displacement)
of the relative position of resonance parameters on the energy and width
$E-\Gamma$ plain. It was shown that the Coulomb shift for bound states is
larger than for resonance states, since bound states are more compact than
resonance states. However, for very narrow resonance states the Coulomb shift
is close to\ the shift of the bound states. This indicates that narrow
resonance states can be treated as compact objects as it has been
demonstrated, for example, in Refs. \cite{2013UkrJPh.58.544V,
2017PhRvC..96c4322V, 2018PhRvC..98b4325V}. Such narrow resonance
states in the three-cluster continuum of $^{9}$Be and $^{9}$B, $^{11}$B and
$^{11}$C \ are the Hoyle-analog states as was shown in Ref.
\cite{2018PhRvC..98b4325V}.

It was also found that the smallest Coulomb shift is observed for broad
resonance states. They are, for example, the 1/2$^{+}$ resonance states in the
three-cluster continuum of $^{9}$Be and $^{9}$B, $^{11}$B and $^{11}$C. As it
was shown in Refs. \cite{2017PhRvC..96c4322V, 2013UkrJPh.58.544V,
2018PhRvC..98b4325V}, these resonance states have very a dispersed
three-cluster structure. Therefore, the Coulomb shifts are equal $R_{C}$=0.429
and $R_{C}$=0.492 MeV, respectively, for these pairs of the mirror nuclei.
There is one pair of resonance states in the two-cluster continuum of the
mirror nuclei $^{8}$Li and $^{8}$B, which also has the smallest Coulomb shift
$R_{C}$=0.304 MeV. This pair of resonance states is the 1$_{3}^{+}$ resonance states.

We have investigated different scenarios of motion of resonance states due to
the Coulomb forces. The dominated scenario for three-cluster systems is when
both energy and width\ of the $R$ nucleus are increased with respect to the
position of corresponding resonance state in the $L$ nucleus. This scenario is
observed for resonance states residing in two- and three-cluster continua. We
also observed a few cases of the second scenario, when the Coulomb interaction
increases energy of resonance state but decreases its width. We have not observed 
the third and fourth scenarios when the Coulomb interaction decreases the energy of resonance states.

And final remark. We have suggested a method how to analyze effects of a specific interaction
(the Coulomb interaction)  on resonance states in two- and three-cluster continuum. 
It is obvious that this method can be applied to study effects of other types of forces 
or different factors on the energies and widths of resonance states in many-channel and/or many-cluster systems.

\bigskip
\section*{Acknowledgement}

Two of the authors (V. Vasilevsky and K. Kat\={o}) are grateful to the members
of the subdepartment of theoretical and nuclear physics from the Physical and
Technical Department, Al-Farabi Kazakh National University, Almaty, Republic
of Kazakhstan, for hospitality and stimulating discussion during their stay at
Al-Farabi Kazakh National University.

This work was supported in part by the Program of Fundamental Research of the
Physics and Astronomy Department of the National Academy of Sciences of
Ukraine (Project No. 0117U000239), by the Ministry of Education and Science of
the Republic of Kazakhstan, Research Grant IRN: AP 05132476 and by JSPS
KAKENHI Grant No. 17K05430.

\end{document}